# Discovery of a novel 1,3,4-oxadiazol-2-one-based NLRP3 inhibitor as a pharmacological agent to mitigate cardiac and metabolic complications in an experimental model of diet-induced metaflammation

Simone Gastaldi [a,1], Carmine Rocca [b,1], Eleonora Gianquinto [a], Maria Concetta Granieri [b], Valentina Boscaro [a], Federica Blua [a], Barbara Rolando [a], Elisabetta Marini [a], Margherita Gallicchio [a], Anna De Bartolo [b], Naomi Romeo [b], Rosa Mazza [b], Francesco Fedele [d,e], Pasquale Pagliaro [c,e,1], Claudia Penna [c,e,1,*], Francesca Spyrakis [a,1,*], Massimo Bertinaria [a,1,*], Tommaso Angelone [b,e,1]

[a] *Department of Drug Science and Technology, University of Turin, 10125 Turin, Italy*

[b] *Cellular and Molecular Cardiovascular Pathophysiology Laboratory, Department of Biology, E. and E.S. (DiBEST), University of Calabria, 87036 Rende, Italy*

[c] *Department of Clinical and Biological Sciences, University of Turin, Turin, Italy*

[d] *Department of Clinical, Internal, Anesthesiology and Cardiovascular Sciences, Sapienza University of Rome, Viale del Policlinico, 155, 00161 Rome, Italy*

[e] *National Institute for Cardiovascular Research (INRC), Bologna, Italy*

[1] These authors equally contributed to this work

## Abstract

Inspired by the recent advancements in understanding the binding mode of sulfonylurea-based NLRP3 inhibitors to the NLRP3 sensor protein, we developed new NLRP3 inhibitors by replacing the central sulfonylurea moiety with different heterocycles. Computational studies evidenced that some of the designed compounds were able to maintain important interaction within the NACHT domain of the target protein similarly to the most active sulfonylurea-based NLRP3 inhibitors. Among the studied compounds, the 1,3,4-oxadiazol-2-one derivative **5** (INF200) showed the most promising results being able to prevent NLRP3-dependent pyroptosis triggered by LPS/ATP and LPS/MSU by 66.3 ± 6.6% and 61.6 ± 11.5% and to reduce IL-1β release (35.5 ± 8.8 % μM) at 10 μM in human macrophages. The selected compound INF200 (20 mg/kg/day) was then tested in an *in vivo* rat model of high-fat diet (HFD)-induced metaflammation to evaluate its beneficial cardiometabolic effects. INF200 significantly counteracted HFD-dependent "anthropometric" changes, improved glucose and lipid profiles, and attenuated systemic inflammation and biomarkers of cardiac dysfunction (particularly BNP). Hemodynamic evaluation on Langendorff model indicate that INF200 limited myocardial damage-dependent ischemia/reperfusion injury (IRI) by improving post-ischemic systolic recovery and attenuating cardiac contracture, infarct size, and LDH release, thus reversing the exacerbation of obesity-associated damage. Mechanistically, in post-ischemic hearts, IFN200 reduced IRI-dependent NLRP3 activation, inflammation, and oxidative stress. These results highlight the potential of the novel NLRP3 inhibitor, INF200, and its ability to reverse the unfavorable cardio-metabolic dysfunction associated with obesity.

.

## Introduction

Obesity is a multifactorial disease related to biological, environmental, lifestyle, and socioeconomic factors. Recently, a great deal of evidence has reported that obesity is associated with the development of cardiovascular diseases (CVD), including hypertension, ventricular hypertrophy and heart failure (HF), and several metabolic disorders, such as insulin resistance and type 2 diabetes. These represent a group of cardio-metabolic diseases (CMD) that are major public health risk factors and important causes of death worldwide. Indeed, fat deposition promotes changes in cardiac structure by altering myocardial function and increasing the risk of developing comorbidities and HF [1, 2]. All the obesity-related complications lead to the activation of a chronic low-grade inflammatory response (i.e. metaflammation) resulting in the production of pro-inflammatory cytokines and activation of specific inflammatory pathways that culminate in CMD [3, 4]. In this context, the nucleotide-binding oligomerization domain (NOD)-like receptor pyrin domain containing 3 (NLRP3) inflammasome complex has emerged as an intracellular machinery responsible for the activation of key intracellular inflammatory pathways [5, 6]. The NLRP3 inflammasome is formed after the activation of the sensor protein (NLRP3), which recruits the apoptosis-related speck-like protein (ASC) and caspase-1. NLRP3 contains three conserved domains: a central nucleotide binding and oligomerization (NACHT), a C-terminal domain rich in leucine (LRR) and an N-terminal effector domain (PYD). ASC and PYD bind to N-terminal motif that recruits the caspase activation domain (CARD), inducing the activation of caspase 1 and promoting the maturation and release of inflammatory cytokines, particularly interleukin (IL)-1β and IL-18, and subsequent inflammatory cell death [3, 4, 6, 7]. In obesity, the NLRP3-dependent inflammation plays a pivotal role in metabolic inflammation and in a variety of metabolic disruptions related to type 2 diabetes mellitus and/or non-alcoholic fatty liver disease (NAFLD) progression, and cardiovascular complications [8, 9]. Specifically, NLRP3 activation contributes to pathological processes leading to onset and progression of several CVDs, including cardiac hypertrophy and maladaptive cardiac fibrosis and HF [10-15]. Noteworthy, NLRP3 inflammasome signaling mediates chronic low-grade inflammation during aging (*i.e., "inflammaging"*), a central and prominent feature associated to HF with preserved ejection fraction (HFpEF) [16]. On the other hand, NLRP3 is activated by reactive oxygen species (ROS) and is involved in myocardial ischemia/reperfusion injury (IRI), as well as in maladaptive cardiac response during hypercaloric regimens, diabetic cardiomyopathy and HF secondary to obesity [17-20].

Although a close interaction between NLRP3, metabolism and related cardiometabolic disorders has been widely observed, the precise participation of NLRP3 in the mechanisms by which myocardial IRI induces inflammation during obesity and metabolic dysfunction requires further investigations. In particular, it is not clear how to obtain the best cardiovascular protection by inhibiting NLRP3. Therefore, improving the knowledge on the molecular mechanisms of NLRP3 inflammasome activation and exploring the effect of its inhibition is of great relevance to attenuate metaflammation, and to treat/prevent CMD. Based on these considerations, we designed and synthetized a novel 1,3,4-oxadiazol-2-one-based NLRP3 inhibitor, INF200 (compound **5**), and investigated its effectiveness in mitigating myocardial and systemic alteration in a rat model of high-fat diet (HFD)-induced metaflammation.

## 2. Results and discussion

*2.1. Design of NLRP3 inhibitors and docking studies*

In the last decade, the investigation for NLRP3 inhibitors raised a great interest in academia and industry, leading to the discovery of both reversible and covalent NLRP3 inhibitors [21-26]. Different chemical scaffolds have been used to design NLRP3 inhibitors, the most effective up to now being di-substituted sulfonylurea typical of glyburide [27]. Notable examples are MCC950 (also known as CRID3) [28] and NP3-146 (Figure 1) [29].

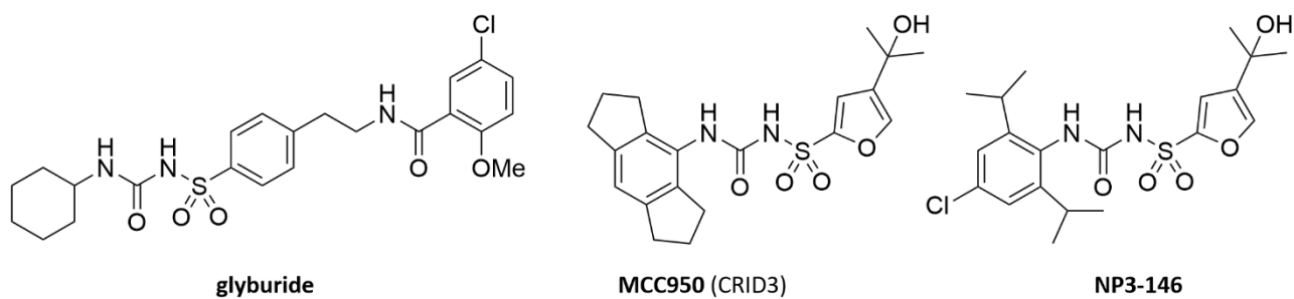

**Figure 1.** Structure of glyburide and of sulfonylurea-derived NLRP3 inhibitors MCC950 and NP3-146.

Sulfonylurea-based inhibitors bind to an allosteric pocket in the NACHT domain of NLRP3, at the interface of HD1, HD2, NBD, FISNA and WHD subdomains and adjacent to the nucleotide binding site, as shown in Figure 2A [29]. This allosteric pocket is shaped by residues belonging to different subdomains of the NACHT domain and is only apparent in the inactive, ADP-bound conformation. Indeed, a recently published Cryo-EM structure [30] shows that, upon activation, HD1, HD2, NBD, FISNA and WHD subdomains undergo a massive conformational change, which requires a 85.4° rigid body rotation of the FISNA-NBD-HD1 module with respect to the WHD-HD2-LRR module, on an axis located between subdomains HD1 and WHD (Figure 2B). Therefore, in the active, ATP-bound form of NLRP3, the allosteric pocket cannot be detected. The inhibitory activity of sulfonylurea-based inhibitors derives from their ability to glue together the HD1, HD2, NBD and WHD subdomains constituting the allosteric pocket, eventually avoiding the activation of NLRP3 NACHT domain.

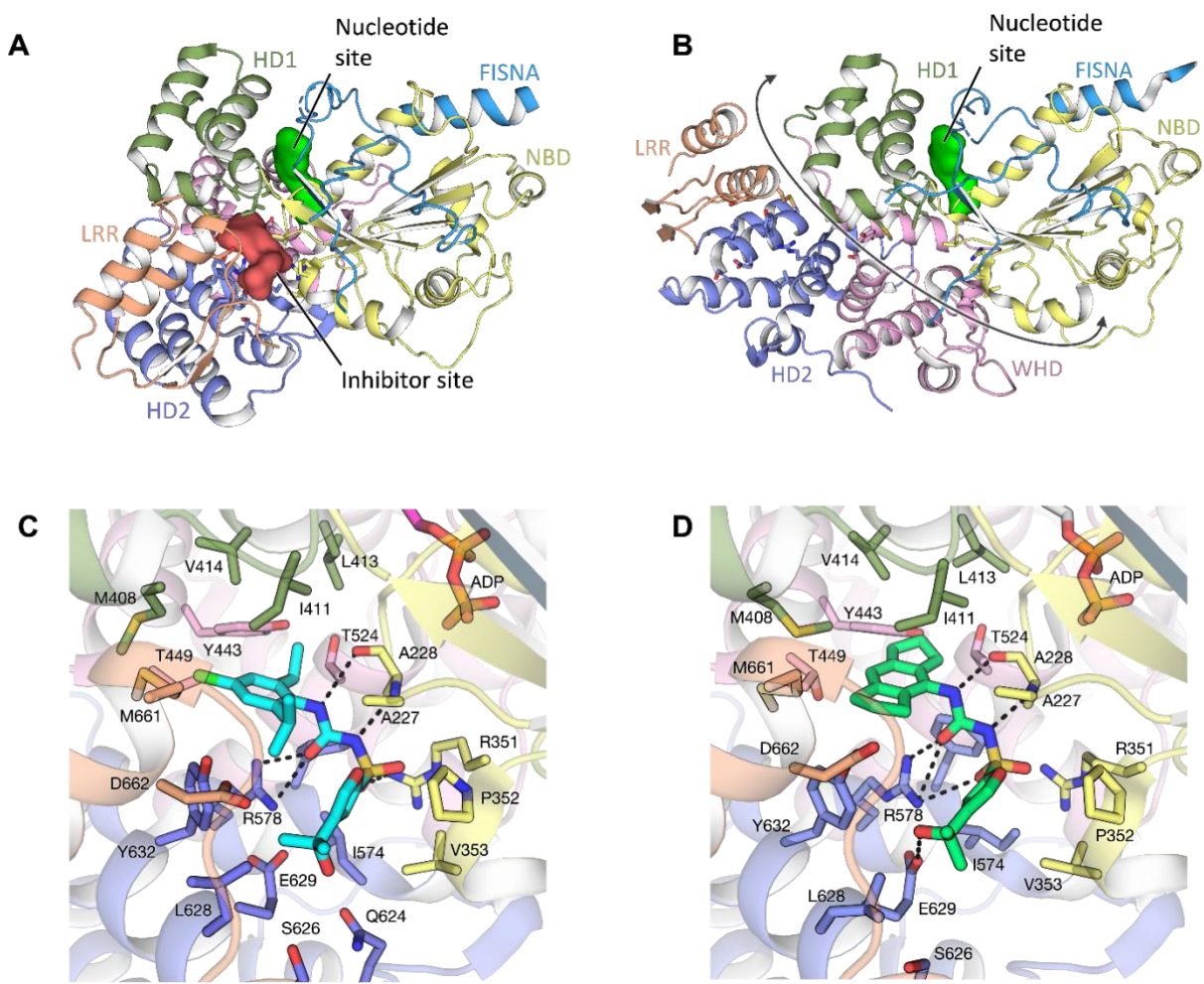

**Figure 2.** (A) Inactive NLRP3 NACHT domain structure, subdomains, and inhibitor (red) and nucleotide (green) binding sites (PDB ID: 7PZC). Residues surrounding the inhibitor pocket are shown in sticks. (B) Active NLRP3 NACHT domain structure, subdomains and nucleotide binding site (PDB ID: 8EJ4). Residues that surround the inhibitor pocket in the inactive NLRP3 NACHT are shown in sticks. (C) Interactions established by NP3-146 in PDB ID: 7ALV [29] and (D) MCC950 in PDB ID: 7PZC [31] into the allosteric NACHT domain pocket of NLRP3. Residues around the inhibitor (NP3-146, cyan; MCC950, orange) are reported as sticks and colored according to the subdomain they belong to, protein is shown as light blue cartoon, distances and angles compatible with hydrogen bonds are highlighted with a black dashed line.

A further crystal structure of the NLRP3 inhibitor NP3-146 in the NACHT domain of NLRP3 has been resolved in late 2021 [29] (Figure 2C), while the Cryo-EM structure of MCC950 bound to NLRP3 has been released in 2022 [31]. These important findings highlighted the binding mode of sulfonylurea-based inhibitors. In Figure 2C, the crystallographic pose of NP3-146 in the NACHT domain of NLRP3 shows that the 2,6-di-isopropyl-4-chlorophenyl moiety is accommodated in a large and mostly lipophilic pocket lined by the residues

Ile230, Ala228, Met408, Phe410, Ile411, Leu413, Val414, Val442, Tyr443, Thr449, Thr524, Phe579, Tyr632 and Met661. In particular, the phenyl ring is sandwiched between residues Ile411 and Arg578. The sulfonylurea core establishes a well-defined network of hydrogen bonds with the backbone of Ala228 and the sidechain of Arg578 and Arg351. Lastly, the 2-hydroxypropan-2-yl-furyl points to the solvent, possibly stabilized by a hydrogen bond with Gln624. The same binding mode was recorded for MCC950 (Figure 2D) in the Cryo-EM structure of inactivated NLRP3.

Based on our previous experience in the development of NLRP3 inhibitors [32-34], we explored the possibility of designing new non-sulfonylurea-based NLRP3 inhibitors. The interaction of the latters with NLRP3 can be generally schematized in three blocks (Figure 3): (i) the lipophilic tail, represented by the *s*-hexaydroindacene in MCC950 or the 4-chloro-2,6-diisopropylbenzene ring in NP3-146 that fits into a hydrophobic pocket; (ii) the sulfonylurea moiety establishes a network of hydrogen bonds with Ala228 and Arg578 of the HD2 subdomain; (iii) the sulfonamide oxygen binds to the positively charged Arg351 of the NBD subdomain. The terminal substituted furanyl moiety in MCC9590 is not essential as it has been extensively modulated by using either aromatic and heteroaromatic rings appropriately substituted to improve physicochemical and pharmacokinetic properties of this class of compounds [27].

Here, we planned to maintain the three mentioned key blocks of interactions by replacing the sulfonylurea-based central core with different heterocycles, while performing limited structural modulations to the lipophilic moiety and terminal polar group. In particular, we used a substituted benzene ring as the lipophilic moiety and a carboxylic acid functionality, or its ester prodrug, as polar group (Figure 4), and linked the three blocks using spacers endowed with different size and flexibility. The spacer chemical nature and length were varied to maintain the interaction with Arg351, as its flexibility can be hardly predicted. The synthetic accessibility of the differently substituted heterocycles was also considered in the design of new compounds.

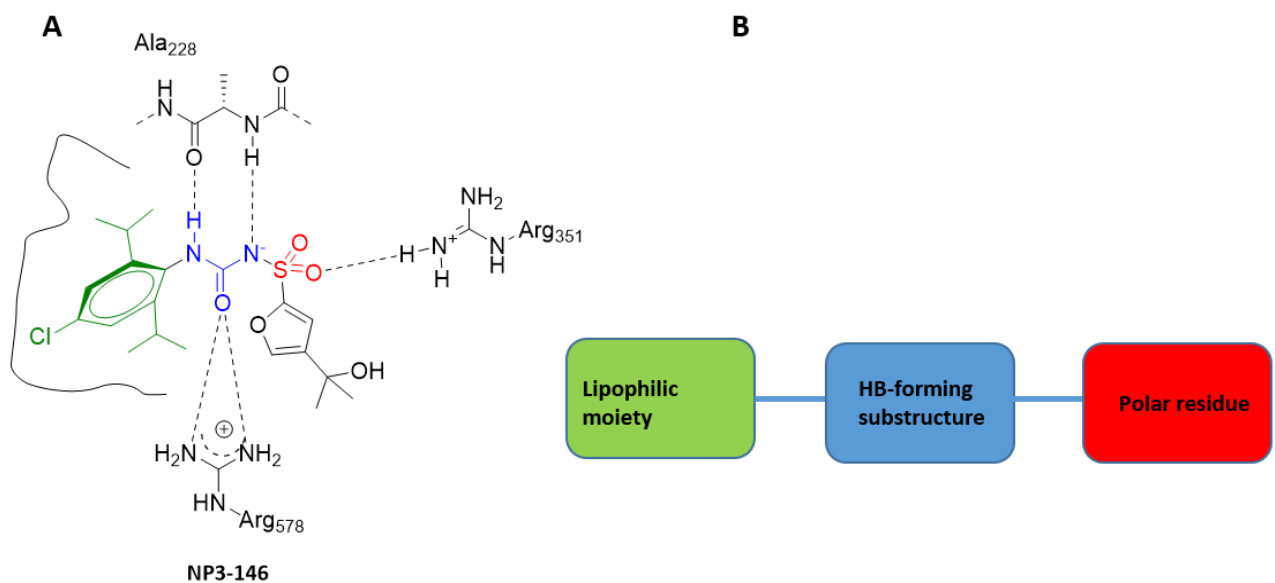

**Figure 3.** A) Schematic representation of the interactions established by NP3-146 into NLRP3 NACHT domain allosteric pocket. The interaction formed by the lipophilic moiety (green), hydrogen bond-forming group (blue) and polar residue (red) are color-coded. B) General structure of the designed compounds.

Docking studies assisted the design process, to get insights on the possible binding mode of compounds in the NACHT allosteric pocket and test whether they could maintain the key interactions established by sulfonylurea-based inhibitors. Synthesized compounds were classified into three series based on their central heterocyclic core: the 1,2,4-oxadiazole series (A), 1,3,4-oxadiazol-2-one series (B), and the 1,3,4-thiadiazole series (C) (Figure 4).

A

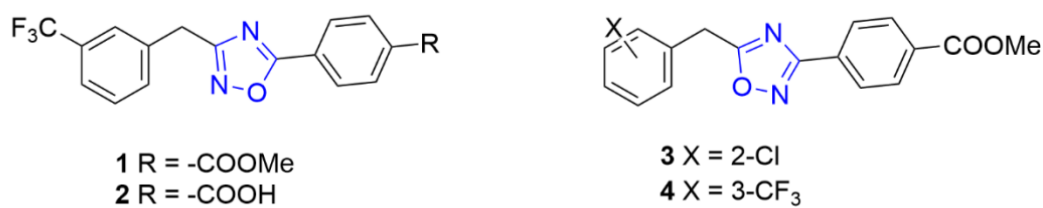

B

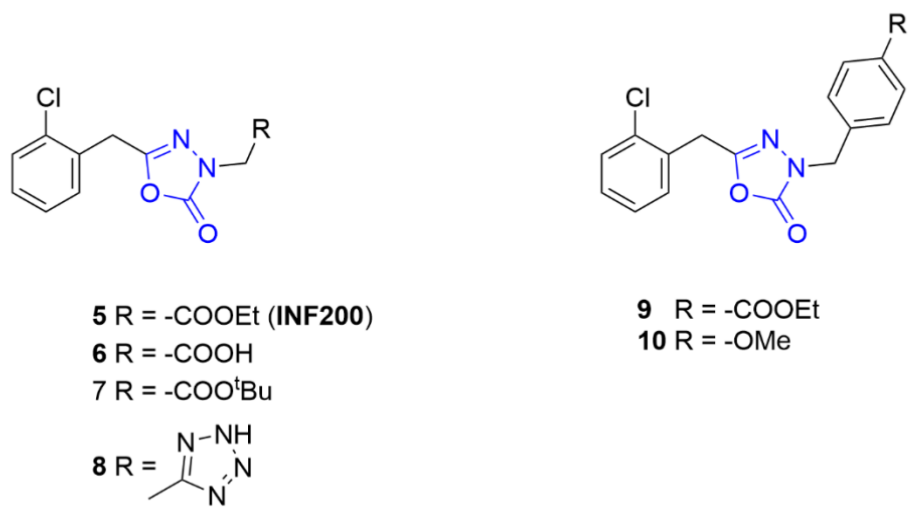

C

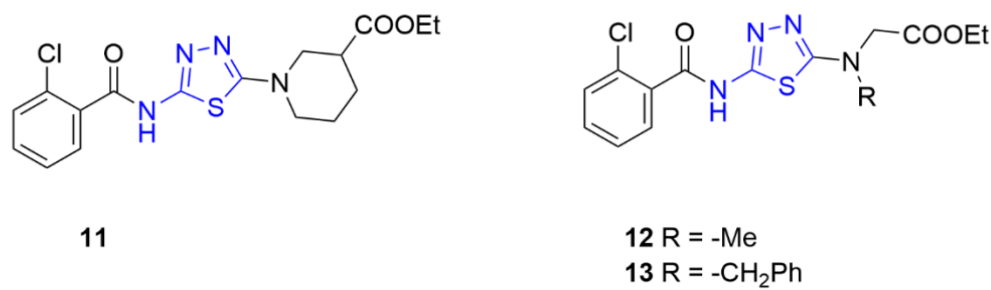

**Figure 4.** Structure of the designed compounds, divided in three series according to their central core (blue): (A) 1,2,4-oxadiazole series; (B) 1,3,4-oxadiazol-2-one series; (C) 1,3,4-thiadiazole series.

**1,2,4-oxadiazole series.** In the first series of compounds, the 1,2,4-oxadiazole ring was employed as the central core (Figure 4A). The lipophilic substituent was introduced in position 3 of the ring (compounds **1** and **2**) or in position 5 (compounds **3** and **4**), respectively. Compounds **1** and **4** are two regioisomers characterized by the spatial inversion of a nitrogen and an oxygen atom in the oxadiazole ring. The polar group is represented by a carboxylic acid group, or by its methyl ester precursor, which should be rapidly hydrolysed in physiological conditions generating the free carboxylic acid, linked to the heterocycle through a phenyl ring.

The docking poses of the 1,2,4-oxadiazole series reflected the rigidity of these compounds, where the main degree of flexibility is given by a methylene spacer, linking the lipophilic tail to the heterocycle. Figure 5A-C reports the top-ranked poses for compounds **2-4**. Overall, two docking modes were recorded, one with the 4-carboxyl-phenyl substituent hydrogen bonding Gln624 and Ser626 (Figure 5A), while the other being more structurally strained and L-shaped, with the 4-carboxyl group establishing a hydrogen bond with the side chain of Arg351 and the backbone of Val353 (Figure 5B and 5C). In both docking modes, the $N^4$ in the oxadiazole ring was oriented towards Arg578, with a distance and an angle compatible with a hydrogen bond. In all docking poses, the phenyl ring was sandwiched between Ile411 and Arg578. 3-trifluoromethyl- and 2-chloro substituents on the phenyl ring filled the same hydrophobic subpockets occupied by isopropyl- and chloro-substituents of NP3-146, respectively. However, the intrinsic molecular rigidity of the 1,2,4-oxadiazole series precluded the generation of a docking pose in which the hallmark hydrogen bonds with Ala228, Arg351 and Arg578 are all fulfilled at the same time. Even if the 1,2,4-oxadiazole series showed a suboptimal pattern of polar interactions in docking studies, we considered that the binding pocket is characterized by high plasticity, especially at the level of Arg351. Such plasticity could not be fully taken into account in docking studies, and we considered that biological testing of this oxadiazole series could be helpful to verify whether these structures can be able to block NLRP3 activation.

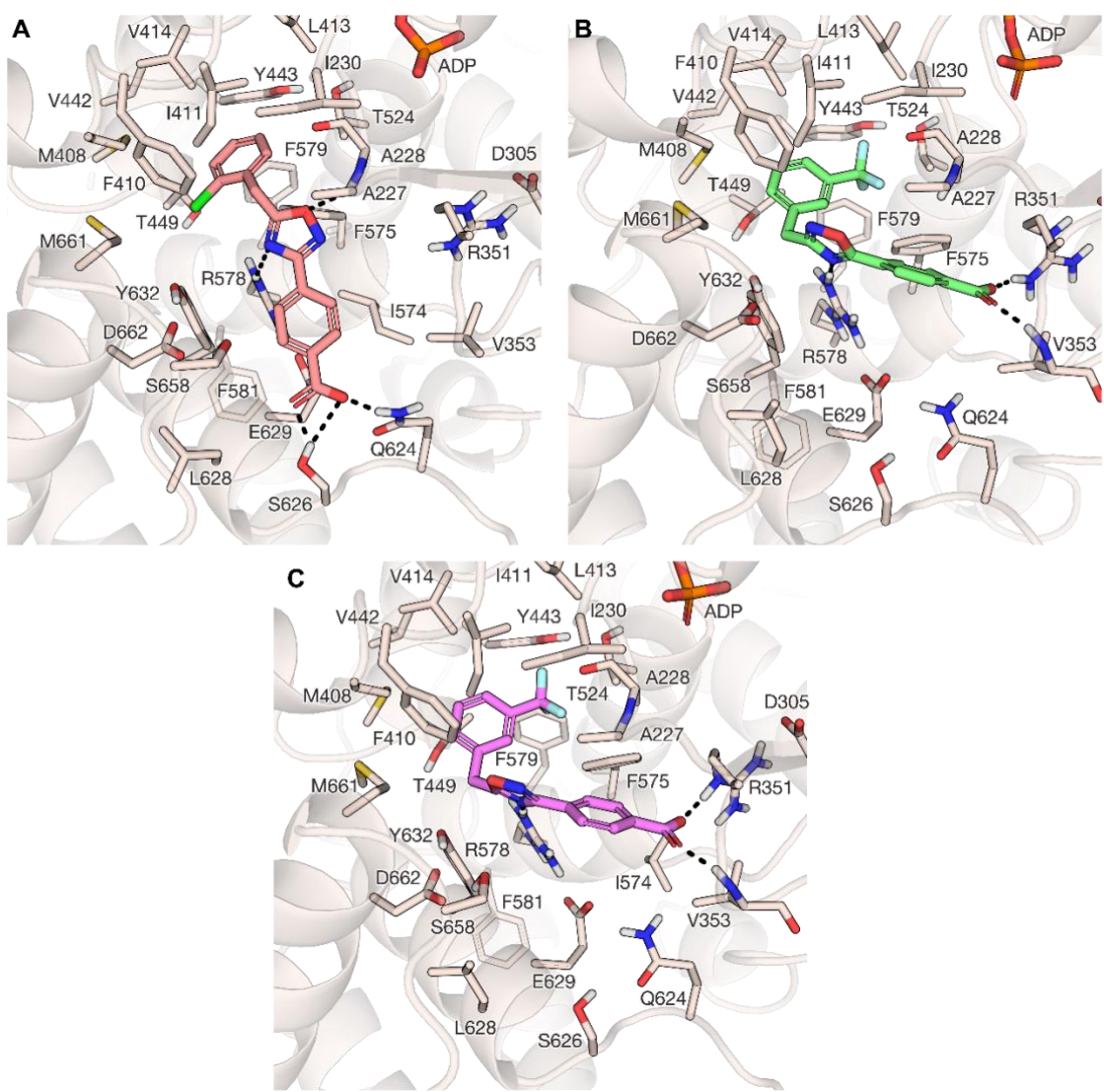

**Figure 5.** Docking poses of the 1,2,4-oxadiazole series in the allosteric site of NLRP3 NACHT domain. The docking pose of representative compounds is reported in a separated panel, where ligand is represented as sticks: (**A**) compound **3**, salmon pink, (**B**) compound **2**, green, and (**C**) compound **4**, violet. Protein is represented as cartoon, binding site residues are depicted as sticks and labelled, putative hydrogen bonds are highlighted by a black dashed line.

**1,3,4-Oxadiazol-2-ones.** The second series of compounds (Figure 4B) bears as central core a 1,3,4-oxadiazol-2-one ring. With respect to the previous series, the use of this heterocycle should additionally allow the interaction of the carbonyl moiety of the 1,3,4-oxadiazol-2-one ring with Arg578. The lipophilic moiety is a 2-chlorobenzyl group, while the polar moiety is represented by a carboxylic acid functionality (compound **6**), or its metabolically labile ethyl ester (compounds **5** and **9**), by a isosteric tetrazole ring (compound **8**), or by a methoxy group (compound **10**). The metabolically stable *tert*-butyl ester (compound **7**) was also designed to verify whether the hydrolysis of ester moiety to the carboxylic acid was an essential requisite for activity. The polar groups mentioned above were linked to $N^3$ of the 1,3,4-oxadiazol-2-one ring through a methylene (compounds **5-8**) or a benzyl group (compounds **9** and **10**).

The docking poses of the metabolically stable form of the 1,3,4-oxadiazol-2-one series showed a conserved binding mode (Figure 6A-D).

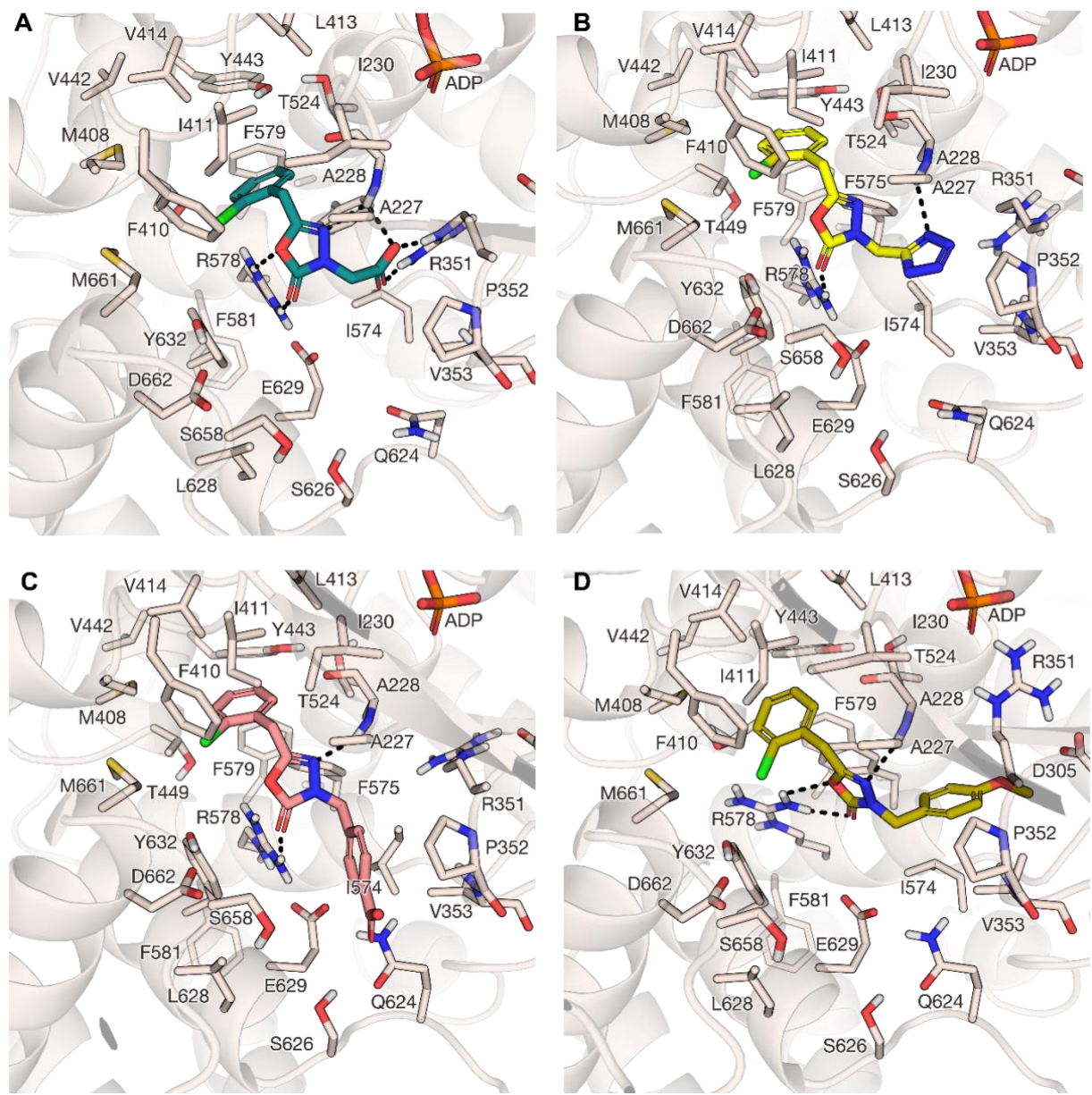

**Figure 6.** Docking poses of the 1,3,4-oxadiazol-2-one series in the allosteric site of NLRP3 NACHT domain. (**A**) Compound **6**, teal (**B**) compound **8**, yellow (**C**) compound **9**, salmon pink**,** and (**D**) compound **10**, dark yellow. Protein is represented as cartoon, binding site residues are depicted as sticks and labelled, putative hydrogen bonds are highlighted by a black dashed line.

In agreement with our hypothesis, the 2-chlorophenyl ring lies in the lipophilic pocket previously defined. The 1,3,4-oxadiazol-2-one heterocycle retained an orientation in which the carbonyl group in $C^2$ hydrogen bonds the side chain of Arg578, while the $N^4$ contacts the backbone of Ala228. The isosteric modification of carboxylic acid on compound **6** (Figure 6A) to a tetrazole group in compound **8** (Figure 6B) was attempted, but this substitution led to an unfavorable shift in the 1,3,4-oxadiazol-2-one core, which moved away from

Ala228, likely because of the increased size of the polar group used. The extension of the linker connecting the 1,3,4-oxadiazol-2-one heterocycle to the polar functionality was incremented inserting a benzyl ring, which led to a pose similar to that obtained for 1,2,4-oxadiazole derivatives of series A and previously described (Figure 6C-D).

**1,3,4-Thiadiazoles.** The last series of compounds (Figure 4C) was designed around the 2,5-diamino-1,3,4-thiadiazole ring. One amino group was substituted with the 2-chlorobenzoyl residue while the second amino group was connected through different linkers to the polar moiety. The 1,3,4-thiadiazole heterocycle is linked to the acidic functionality by a tertiary amine group, which in the case of compound **11** is enclosed in a piperidinyl ring bearing the carboxylate in position 3', thus leading to the existence of (R-) and (S-) stereoisomers. Contrarily, in compound **12** and **13** the tertiary amine linking the thiadiazole ring to the carboxylic acid group is not constrained in a ring, and bears an N-methyl or an N-benzyl substituent (Figure 4).

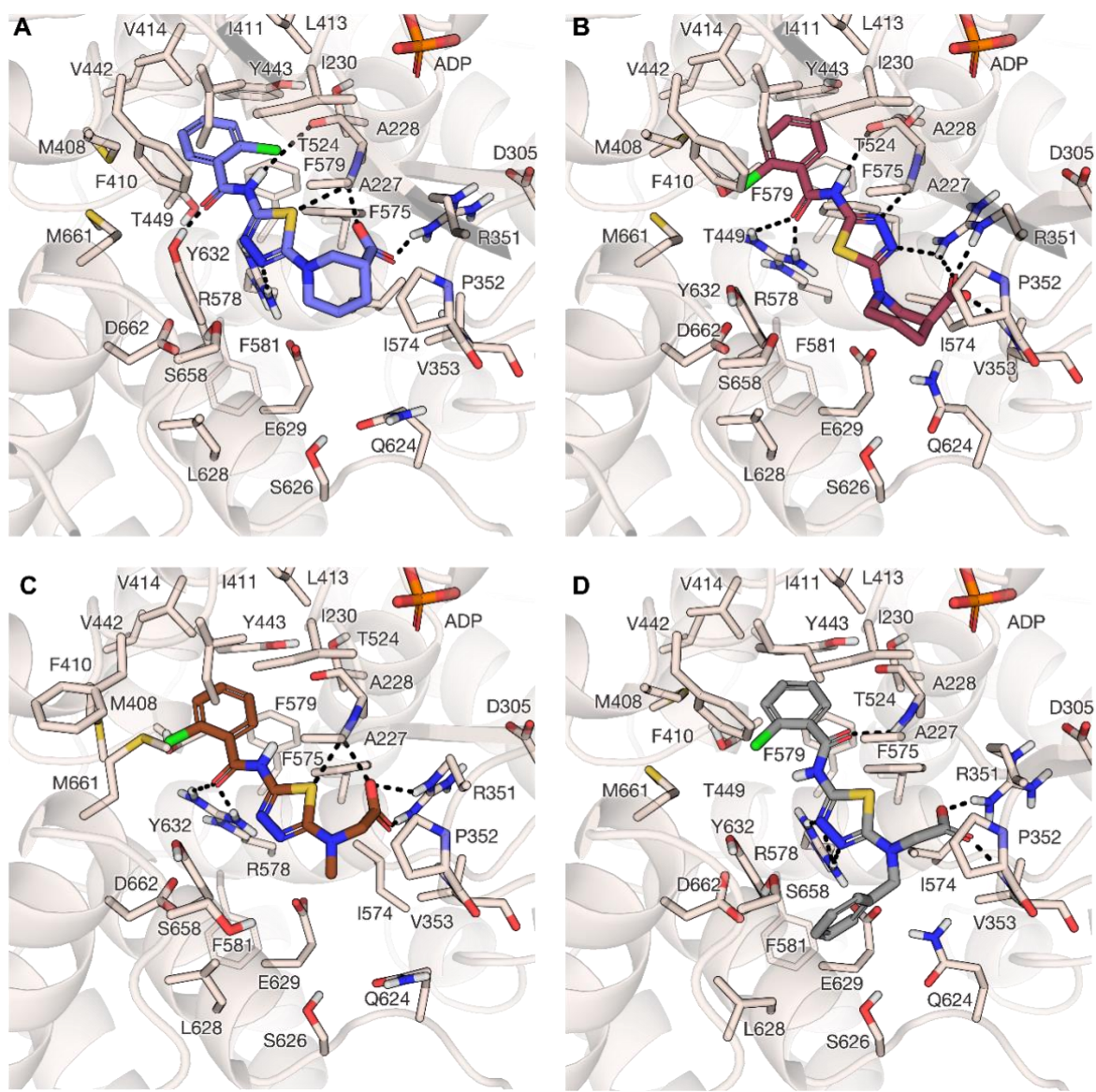

**Figure 7.** Docking poses of the 1,3,4-thiadiazole series in the allosteric site of NLRP3 NACHT domain. (**A**) Compound **11** (R stereoisomer), purple (**B**) compound **11** (S stereoisomer), dark red (**C**) compound **12**, brown**,** and (**D**) compound **13**, grey. Protein is represented as cartoon, binding site residues are depicted as sticks and labelled, putative hydrogen bonds are highlighted by a black dashed line.

The amide linker, added to recover a bidentate hydrogen bond with Ala228, as observed in the NP3-146 X-ray pose, provided the compounds with greater flexibility. Indeed, the interactions established by this group were not conserved in all the poses. In all the cases, the carboxylic acid group established ionic interactions with the side chain of Arg351, mimicking the position and role of the sulfonyl group of NP3-146. Instead, the thiadiazole ring was able to explore several conformations, among which a frequently generated one shows the $S^1$ atom pointing to the backbone of Ala228, while $N^3$ and $N^4$ are oriented toward Arg578 (Figure 7A-D).

*2.2. Chemistry*

The 1,2,4-oxadiazoles were synthesized according to Scheme 1. The synthesis of compounds **1** and **2** required the production of the intermediate trifluoromethyl 4-(N'-hydroxycarbamimidoyl)benzoate (**15**) which was obtained by heating a mixture of commercially available 3-trifluoromethylphenyl acetonitrile (**14**) and hydroxylamine in EtOH for 3 h. The resulting hydroxyimidamide (**15**) was then reacted with methyl-4-formyl benzoate in the presence of a catalytic amount of p-toluenesulfonic acid (PTSA) to afford **1** in low yield (15%). The desired acid derivative **2** was obtained from **1** through hydrolysis using an aqueous solution of LiOH in THF. Since **1** was obtained in low yield using the above reported reaction conditions, a different synthetic pathway was pursued for the synthesis of the 1,2,4-oxadiazole ring in compounds **3** and **4**. In this case, the cyclization was carried out in a two-step procedure. The coupling agent carbonyl diimidazole (CDI) activated the 2-(3-chlorophenyl)acetic acid, which undergoes the hydroxyimidamide (**17**) attack, causing the acid-promoted internal cyclization to afford compound **3**. The same procedure was adopted for the synthesis of **4**, using 2-(3-(trifluoromethyl)phenyl)acetic acid as the coupling partner of the hydroxyimidamide **17** affording the desired derivative **4** in reasonable yields.

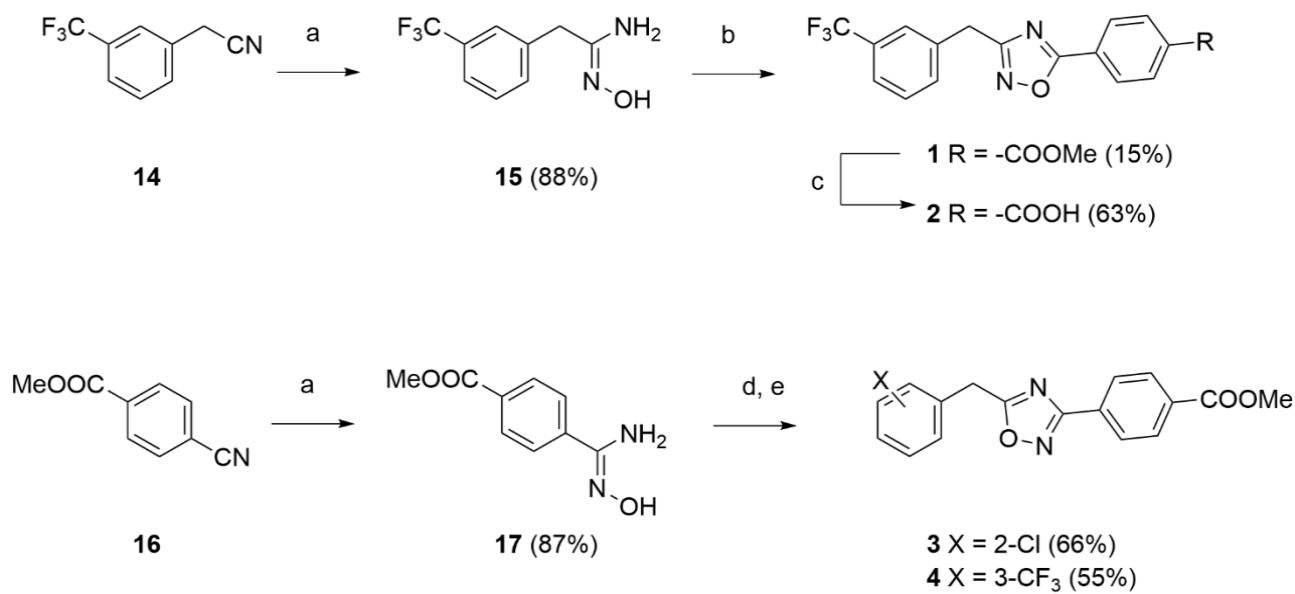

**Scheme 1**. *Reagents and conditions:* (a) hydroxylamine hydrochloride (2.2 eq), TEA (2.3 eq), EtOH 96%, 75 °C, 5 h. (b) methyl 4-formylbenzoate (1.1 eq), PTSA (0.1 eq), dry toluene, reflux, 18 h. (c) LiOH (5 eq), THF, rt, 18 h. (d) appropriate phenyl acetic acid (1.1 eq), CDI (1.1 eq), THF, rt, 3h. (e) Acetic acid, 118 °C, 18 h.

For the synthesis of 1,3,4-oxadiazol-2-ones **5**, **7**, **9**, **10**, **21** (Scheme 2), 2-(chlorophenyl)acetic acid (**18**) was converted into the corresponding hydrazide **19** using hydrazine and CDI as the coupling agent. This intermediate was then cyclized using another equivalent of CDI to obtain 5-(2-chlorobenzyl)-1,3,4-oxadiazol-2(3*H*)-one (**20**). The 1,3,4-oxadiazol-2(3*H*)-one ring was then functionalized with the desired substituents. The presence of an acid hydrogen on the ring in **20** (predicted p$k_a$ of 6.34, calculated using Advanced Chemistry Development ACD/Labs Software V11.02, measured pka = 7.41 [35]) allowed the formation of a new C-N bond by using 1,8-Diazabicyclo[5.4.0]undec-7-ene (DBU)-mediated nucleophilic substitution, leading to compounds **5**, **7**, **9** and **21**. The electrophilic reagents were respectively methyl 2-bromoacetate, *t*-butyl-2-bromoacetate, 2-bromoacetonitrile and ethyl (4-bromomethyl)benzoate. The intermediate **21** was cyclized using the click reaction between the nitrile and sodium azide, in presence of ammonium chloride in DMF to give **8** in high yields. The acid hydrolysis of the *tert*-butyl ester of **7** gave compound **6**. We observed that it was not possible to obtain compound **6** by basic hydrolysis of the ethyl ester derivative **5**, owing to decomposition of the heterocycle under these conditions. To obtain compound **10**, we exploited the acid nature of the NH group in 1,3,4-oxadiazol-2(3*H*)-one ring by using the Mitsunobu reaction of **20** with 4-methoxybenzyl alcohol.

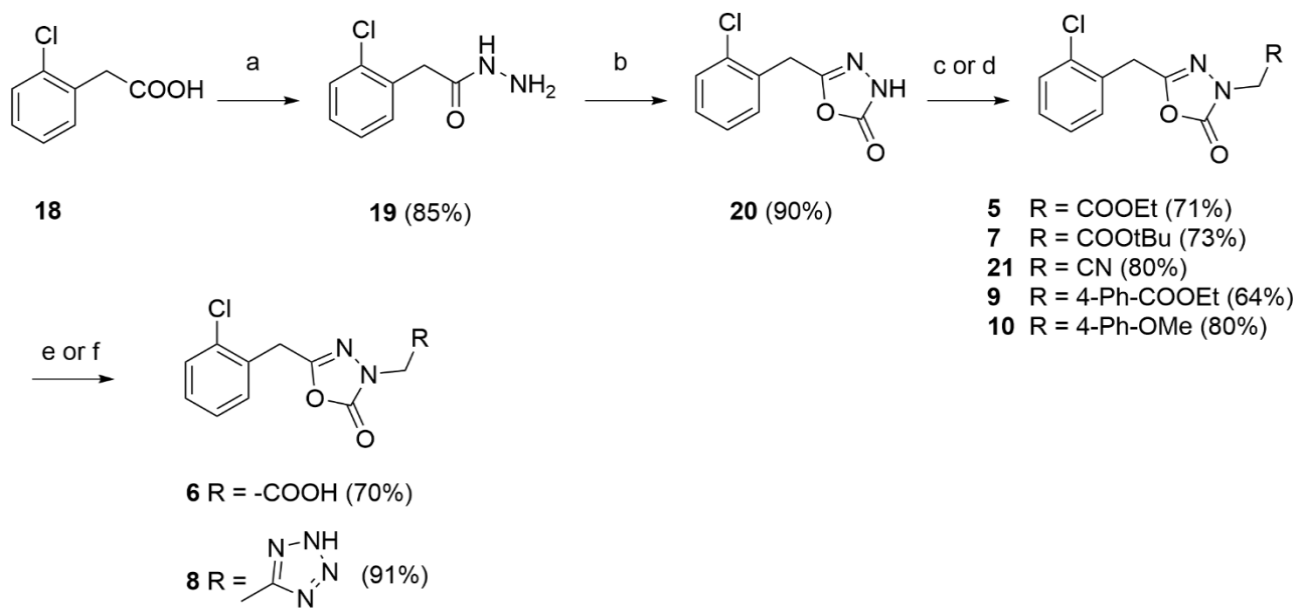

**Scheme 2**. *Reagents and conditions:* (a) CDI (1.1 eq), $NH_2NH_2 \cdot H_2O$ (1.5 eq), THF, rt, 18 h. (b) CDI (1.1 eq), dry THF, rt, 18 h. (c) DBU (1.5 eq), ethyl 2-bromoacetate, or t-Bu 2-bromoacetate, or 2-bromoacetonitrile, or ethyl (4-bromomethyl)benzoate (2 eq), THF, rt, 18 h. (d) DIAD (1.5 eq), 4-methoxybenzyl alcohol (1 eq), $PPh_3$ (1.5 eq), THF, 0 °C to rt, 4 h. (e) **7**, TFA (10 eq), DCM, rt, 18 h. (f) **21**, $NaN_3$ (1.5 eq), $NH_4Cl$ (1 eq), DMF, rt, 2 h.

Finally, the synthesis of the 1,3,4-thiadiazoles was obtained according to Scheme 3. The activated ester **23** was prepared by dicyclohexylcarbodiimide (DCC)-mediated coupling of 2-chlorobenzoic acid (**22**) with N-hydroxysuccinimide (NHS), while the intermediate thiadiazoles **25**-**27** were prepared by nucleophilic aromatic substitution on commercially available 2-bromo-5-amino-1,3,4-thiadiazole (**24**). This reaction was performed by treating **24** with ethyl nipecotate, ethyl N-methylglycinate and ethyl N-benzylglycinate in DMF at 80 °C for 2 h, in presence of DIPEA as the base. By employing this procedure, the substituted thiadiazoles **25**-**27** were obtained in satisfactory yields. The reaction of intermediates **25**-**27** with the activated NHS-ester **23** required forcing conditions as it was necessary to conduct it at 100 °C in DMF to obtain good yields of the desired final compounds **11**-**13**.

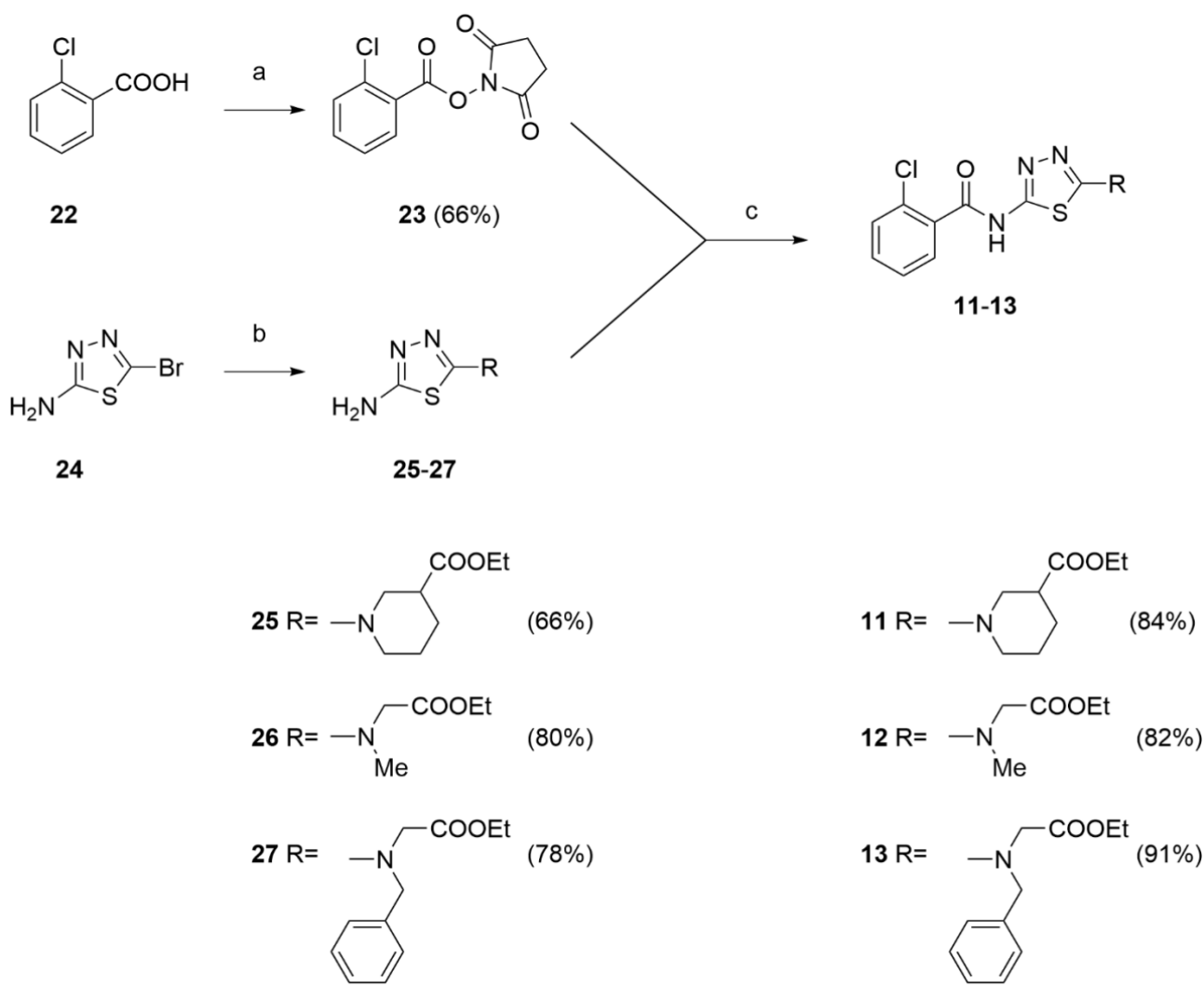

**Scheme 3**. *Reagents and conditions:* (a) DCC (1 eq), NHS (1.5 eq), THF, 0 °C to rt, 18 h. (b) ethyl nipecotate or ethyl N-methylglycinate or ethyl N-benzylglycinate (1.1 eq), DIPEA (5 eq), DMF, 80 °C, 2 h. (c) **23**, amines **25-27** (1 eq), DIPEA (1 eq), DMF, 100 °C, 18 h.

*2.3. In vitro biological evaluations*

The synthesized compounds were characterized for their ability to inhibit NLRP3-dependent pyroptosis in differentiated THP-1 cells. Briefly, THP-1 cells were plated and differentiated into macrophages by treatment with phorbol myristate acetate (PMA; 50 nM; 24 h). Differentiated cells were primed with lipopolysaccharide (LPS; 10 μg/mL; 4 h) in serum-free medium and then treated with either vehicle alone or test compound (10 μM; 1 h). The cell death was triggered with ATP (5 mM), and the pyroptotic cell death evaluated after 1.5 h by measuring the LDH released in the cell supernatants according to the published procedure [36]. The obtained data, expressed as a percent of pyroptosis decrease *vs* vehicle alone, are reported in Table 1. For those compounds showing > 25 % inhibition of pyroptotic cell death at 10 μM, the % inhibition of IL-1β release in THP-1 cells was also evaluated via an ELISA assay (Table 1). Since all in silico simulations were performed on the PDB ID 7alv, the X-ray structure of NLRP3 co-crystallyzed with NP3-146, the activity of the newly synthesized compounds was compared to that of the reference compound NP3-146.

Finally, the cytotoxicity exerted by test compounds over 72 h was evaluated using the MTT assay after treatment of THP-1 cells with increasing concentration of test compounds (0-1 – 100 µM; Figure S1); the results, expressed as $TC_{50}$, are collected in Table 1.

**Table 1.** Inhibitory effect of synthesized compounds on pyroptotic cell death, IL-1β release in differentiated THP-1 cells and cytotoxicity in THP-1 cells.

| Compound | Pyroptosis decrease[a] % inhibition ± SEM at 10 µM | IL-1β release[b] % inhibition ± SEM at 10 µM | Cytotoxicity[c] $TC_{50}$ (µM) |
|---|---|---|---|
| 1 | 46.3 ± 17.3[d] | < 10 | > 100 |
| 2 | 14.2 ± 2.1 | NT | 27.3 ± 1.3 |
| 3 | < 10 | NT | > 100 |
| 4 | < 10 | NT | 94.5 ± 1.1 |
| 5 | 66.3 ± 6.6[e] | 35.5 ± 8.1[d] | 76.5 ± 1.2 |
| 6 | 45.9 ± 8.4[d] | 30.3 ± 14.6[d] | 94.6 ± 1.2 |
| 7 | < 10 | NT | 98.9 ± 15.0 |
| 8 | 27.6 ± 9.5 | <10 | > 100 |
| 9 | 17.7 ± 6.4 | NT | > 100 |
| 10 | 11.4 ± 3.8 | NT | 64.8 ± 5.6 |
| 11 | 39.7 ± 13.3 | 22.4 ± 9.5 | 79.7 ± 7.0 |
| 12 | 25.7 ± 5.8 | 20.7 ± 12.6 | 96.1 ± 4.0 |
| 13 | 60.0 ± 13.2[e] | 49.0 ± 12.4[e] | 91.6 ± 14.5 |
| NP3-146 | 56.1 ± 2.5[e] | 54.7 ± 7.2[e] | 92.5 ± 4.1 |

[a] Pyroptosis of differentiated THP-1 cells was triggered using LPS/ATP. Data are reported as the % inhibition of pyroptosis of cells treated with 10 µM conc. of test compound *vs* vehicle-treated cells. Data are the mean ± SEM of three to five experiments run in triplicate.
[b] IL-1β inhibition was measured in the cell supernatants form the same experiments. Data are reported as % inhibition ± SEM of three to five experiments run in triplicate.
[c] Cytotoxicity was determined after 72 h treatment of THP-1 cells with increasing conc. (0.1–100 μM) of test compounds. Data are reported as $TC_{50}$ ± SEM of three experiments.
[d] $p < 0.05$ *vs* vehicle treated cells;
[e] $p < 0.01$ *vs* vehicle treated cells;
NT = not tested.

Among the synthesized 1,2,4-oxadiazole derivatives (**1**–**4**), only compound **1** showed a relevant antipyroptotic activity at 10 µM (46.3 ± 17.3%), however, it was not able to reduce IL-1β release in the same cellular system. The hydrolysis of the terminal methyl ester of **1** to afford the corresponding free acid **2** resulted in a significant loss of antipyroptotic activity (14.2 ± 2.1%). Both compounds **3** and **4**, in which the position of the lipophilic moiety and the polar residue was reversed around the 1,2,4-oxadiazole ring, were found to be inactive. This proved in agreement with observations made during the design process, where the rigidity of the 1,2,4-oxadiazole compounds resulted in suboptimal docking poses.

The compounds from the second series (**5-10**), bearing the 1,3,4-oxadiazol-2-one ring as the central core, displayed interesting results (Table 1). Compound **6**, which carries the acetic acid residue as its terminal polar group, showed good ability to inhibit NLRP3-dependent pyroptotic cell death and IL-1β release (Table 1). The corresponding ethyl ester prodrug **5** showed an increase in the antipyroptotic activity with respect to the parent acid **6** (66.3 % *vs* 45.9 % inhibition at 10 µM, see Table 1). This is not surprising since this simple modulation increases the lipophilicity at physiological pH; shake flask experiments have shown that the *n*-octanol/water distribution coefficient at physiological pH of **5** (log $D^{7.4}$ = 2.03 ±0.09) is several logarithmic units greater than that of **6** (log $D^{7.4}$ = -1.66 ±0.06), thereby favoring the cell permeability, which reflects in an increased activity. This behavior was previously observed both by our group, working with acrylate derivatives, and by researchers at Nodthera working with MCC950 derivatives [32, 37]. We hypothesize that, after the ethyl ester penetrate into the cell, it is easily hydrolyzed by the cellular esterases to produce the free carboxylic acid, which could establish more productive interactions with the putative binding site into the NACHT domain of NLRP3. In agreement with this observation, the *tert*-butyl ester derivative **7**, which is more lipophilic (log $D^{7.4}$ = 3.1 ±0.1) but stable to esterases, proved inactive. Moreover, the tetrazole derivative **8** (log $D^{7.4}$ = -1.09 ±0.09), sharing similar physico-chemical properties with **6**, showed a slightly reduced antipyroptotic activity, likely because of the less favorable pose in the allosteric site. Compounds **5** and **6** were also able to hamper IL-1β production in LPS/ATP-stimulated cells with an inhibition in the 30 – 35% range at 10 µM (Table 1) and no relevant cytotoxic effect at the active concentration ($TC_{50}$= 76.5 and 94.6 µM). The insertion of a phenyl ring between the heterocycle and the carboxy group yielded the ester derivative **9**, with reduced potency (pyroptosis reduction: 17.7 ± 6.4%) compared to the hit compound **5**. The replacement of the carboxylic group in **9** with a polar-non-hydrolyzable methoxy group (compound **10**) almost abrogated the antipyroptotic activity. In agreement with what observed for the 1,2,4-oxadiazole compounds, also in this case compounds with longer and bulkier spacers (**8**, **9**, **10**) between the heterocycle and the acid function showed a decreased NLRP3 inhibition. As already shown by docking studies, we speculate that the presence of a phenyl or benzyl spacer may drastically change the binding mode and might preclude hydrogen bond formation with Arg351.

Compound **5** (i.e., the ethyl ester prodrug of compound **6**) showed the best inhibitory activity among the 1,3,4-oxadiazole-2-one series. Interestingly, compound **6** was endowed with the most conserved docking pose, retaining most of the key polar contacts established by sulfonylurea-based inhibitors in the NACHT pocket.

Compounds **11-13**, built around the 1,3,4-thiadiazole ring, gave interesting results. All of them were able to prevent pyroptosis of differentiated THP-1 cells treated with LPS/ATP at 10 µM concentration with negligible cytotoxicity. Even if the synthesis of further compounds is needed to draw an accurate SAR for this series of 1,3,4-thiadiazole derivatives, the preliminary data obtained (Table 1) suggest that the amino group in position 5 of the ring can be disubstituted thereby modulating both lipophilicity and flexibility of this series of compounds. In this series, the N-benzyl-substituted compound **13** proved the most active with antipyroptotic and anti-IL-1β activities similar to that shown by the oxadiazol-2-one derivative **5**. As already observed during the design phase, all the 1,3,4-thiadiazole compounds displayed promising docking poses where the 2-clorophenyl ring occupied the lipophilic pocket and a rich pattern of polar interactions involving key residues Ala228, Arg351 and Arg578 was recorded. Nevertheless, among the top-ranked docking poses we noticed that the 1,3,4-thiadiazole compounds had more variation compared to compound **5**, which in contrast consistently docked in the same orientation.

From this screening, the 1,3,4-oxadiazol-2-one was identified as promising replacement for the sulfonylurea core in the design of new NLRP3 inhibitors, therefore compound **5** was selected for further studies.

First, we checked and confirmed the activity of **5** by using sodium monourate (MSU; 200μg/mL), a stimulus able to activate the NLRP3-inflammasome with a different mechanism compared to ATP [38-40]. Indeed, compound **5** was able to inhibit NLRP3-dependent pyroptosis by 61.6 ± 11.5 % at 10 μM. The release of TNF-α from differentiated-THP-1 was measured after treatment with compounds **5** and **6** at 10 μM (Figure S2). In this experiment, TNF-α release was not significantly reduced evidencing an absence of direct effects on the NF-kB signaling.

Finally, we performed concentration-response experiments that confirmed compounds **5** and **6** able to inhibit pyroptosis and IL-1β release triggered by LPS/ATP in a concentration-dependent manner (Figure 8); the calculated $IC_{50}$ values for pyroptosis prevention and IL-1 β inhibition were 1.55 ± 2.20 μM and 16,61 ± 2.57 μM for **5** and 1.85 ± 2.68 μM and 32.55 ± 5.04 μM for **6**, respectively

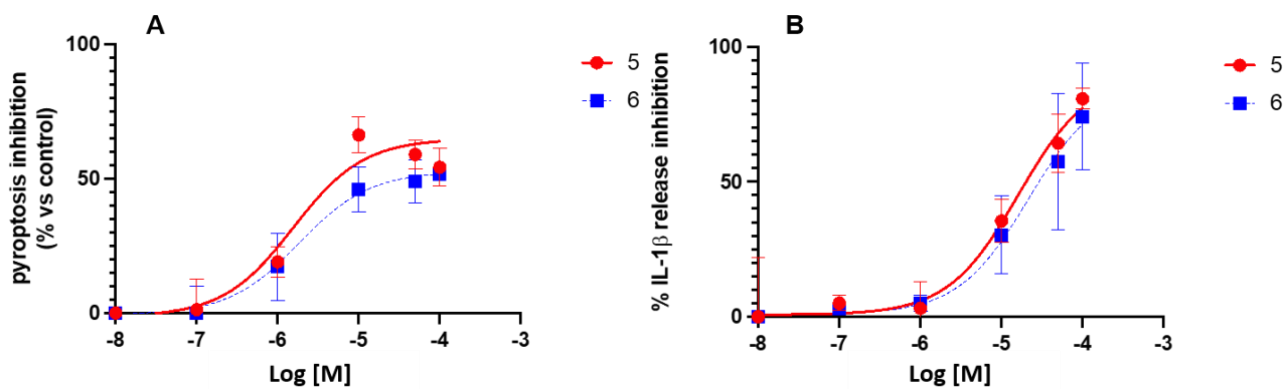

**Figure 8.** Concentration-response curves for compounds **5** and **6** in differentiated THP-1 cells. A) Prevention of pyroptotic cell death triggered with LPS/ATP by increasing concentration (0.01 – 100 μM) of compounds **5** and **6**; B) inhibition of IL-1β release triggered with LPS/ATP by increasing concentration (0.01 – 100 μM) of compounds **5** and **6**. Data are the mean ± SEM of three independent experiments.

*2.4. Molecular dynamics of compound 6 in NLRP3 NACHT domain*

To rationalize the activity of **6** (the expected intracellular metabolite of **5**) compared to the NP3-146 reference compound, a set of 5 unbiased molecular dynamics (MD) replicas (300 ns each) were run for two systems, (i) the reference NP3-146 inhibitor bound to the NLRP3 NACHT domain (PDB ID: 7ALV) and (ii) the docking pose of compound **6** in the NLRP3 NACHT domain. To compare the pose stability of compound **6** with the reference NP3-146 inhibitor, the root mean square deviation (RMSD) of the ligands along all replicas was evaluated (Figure S3). While NP3-146 remained stably bound to the allosteric pocket in its crystallographic pose in all replicas, the conformation of compound **6** changed with respect to the docking pose. As shown by the atomic root mean square fluctuation (RMSF) in Figure S4, the very low variability in the pose of NP3-146 was due to the rotation of the 2-hydroxy-propanoyl substituent, while the rest of the molecule remained very stably anchored to the pocket. In contrast, compound **6** showed a higher atomic RMSF in average, where fluctuations involved both the carboxylic acid group and the oxadiazol-2-one core. This difference could also be spotted when monitoring the RMSF of key residue side chains in the binding pocket. As reported in Figure S5, the average RMSF across replicas for binding site residues Ala228, Arg351, Pro352, Val414, Phe575, Arg578 and Phe579 was higher when NACHT was bound to compound **6**, compared to the same system bound to NP3-146. To further investigate the differences between NP3-146 and compound **6** in the NACHT allosteric pocket, we evaluated the pattern of hydrogen bonds established by NP3-146 and compound **6** along MD simulations. Figure S6A-B shows the occupancy of hydrogen bonds (i.e., percentage of frames calculated on a time window, in which a hydrogen bond donor-acceptor atomic pair has a distance and an angle compatible with a hydrogen bond, see Methods) along MD replicas. NP3-146 established hydrogen bonds with both donor and acceptor groups, while compound **6** could only behave as H-bond acceptor. However, in Figure S6C-D the polar interactions established by NP3-146 H-bond donor groups were relatively transient compared to the H-bond acceptor contributions. The carboxyl group in compound **6** retained hydrogen bonds with relevant polar anchors such as Arg351, Ala228 and Arg578, even if less persistently than NP3-146. To investigate the origin of NP3-146 stability with respect to compound **6**, we observed that the bulky 4-chloro-2-diisopropyl-phenyl moiety in NP3-146 inhibitor filled well the lipophilic cavity of the allosteric pocket, while the chlorophenyl substituent in compound **6** was too small to completely fill the lipophilic cavity and could not shield water molecules from entering in the pocket. Indeed, considering the hydration of the allosteric pocket in all replicas (Figure S7-8), compound **6**, being smaller in size than the reference inhibitor NP3-146, allowed the entering of water molecules from the bulk, which contributed to the conformational variability observed in the RMSD analysis. Further modulations of oxadiazol-2-one derivatives may involve the insertion of a bulkier lipophilic substituent that could satisfy the requirements of the lipophilic subpocket of the NACHT allosteric site. In particular, the use of the 1,2,3,5,6,7-hexahydro-*s*-indacene or the 4-chloro-2,6-diisopropylphenyl moieties, as well as other bulky substituents, could be a modulation to achieve this aim.

*2.5. Stability studies*

We then verified the relative stability of the 1,3,4-oxadiazol-2-one ring and the rate of the ethyl ester group hydrolysis. Compound **5** was incubated in PBS at pH 7.4 at 37 °C and the mixture was analyzed by HPLC at different time intervals over 48 h. In these conditions, **5** was slowly hydrolyzed with an estimated half-time greater than 48 h to produce the free acid **6**, as the only reaction product (Figure S9). The experiment conducted in human serum gave similar results, with the acid **6** identified as the only reaction product. However, in this case a half-time of 4 h was calculated for the hydrolysis of **5** to **6** (Figure 9A). Finally, compound **5** was incubated in THP-1 cell lysate, showing a rapid hydrolysis to **6** with an estimated half-time of 65 min (Figure 9B). As expected, the *tert*-butyl ester **7** was stable in human serum (Figure S10); only traces (< 3%) of the free acid **6** were detected after 24 h at 37 °C. Compound **6**, the expected intracellular metabolite of **5**, was completely stable for 24 h in all the studied conditions (data not shown).

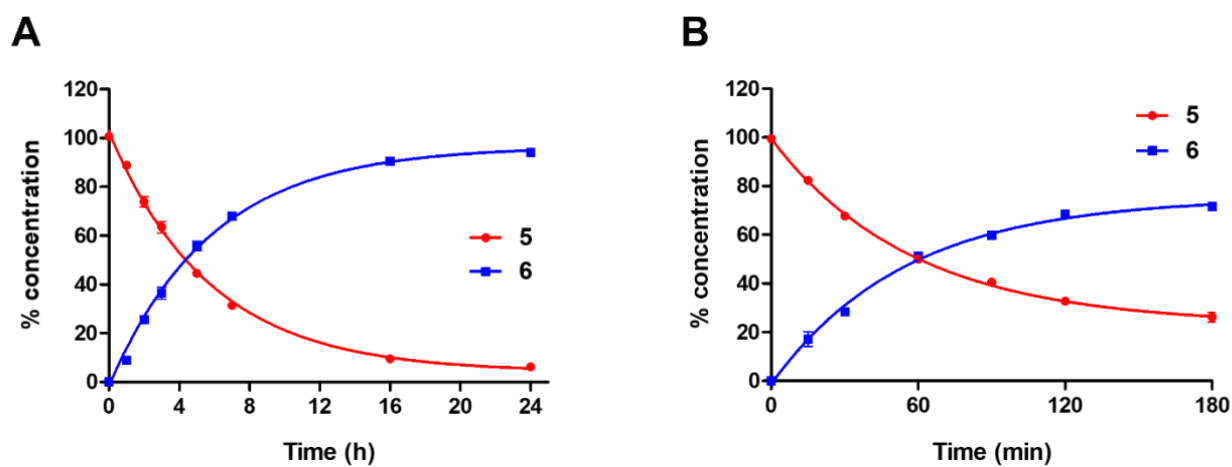

**Figure 9.** Stability of compound **5** under different conditions. A) Incubation of **5** at 37 °C in human serum. B) Incubation of **5** at 37 °C in THP-1 cells lysate. Compound **5** and its degradation product **6** were quantified by RP-HPLC analysis (quantification was done using calibration with external standards of **5** and **6**). Results are the mean ± SEM of three independent experiments.

The obtained data demonstrated that compound **5**, hereafter named INF200, was well suited for *in vivo* studies in a model of HFD-induced metaflammation.

*2.6. In vivo study*

*2.6.1. INF200 improved «anthropometric» alterations induced by HFD regimen.*

For *in vivo* intraperitoneal administration of compound **5** we used a saline solution containing 2% DMSO and 5% Cremophor-EL; the obtained solution of compound **5** was stable for more than 7 days with no observed decomposition or precipitation when stored at 4 °C and 25 °C (Figure S11).

In assessing the *in vivo* beneficial action of INF200 against HFD-induced metaflammation, we firstly investigated its effect on "anthropometric" variables in a rat model. Figure 10A shows that rats fed with HFD alone exhibited a significant body weight increase compared with the control group, *i.e.* rats fed with standard diet (SD). INF200, i.p. administered (20 mg/kg/day) in the last 4 weeks of diet regimen mitigated HFD-induced body weight increase, although this effect was not statistically significant. After 12 weeks, we evaluated the fat gain as a main determinant of obesity development. In particular, abdominal, perirenal, epididymal and retroperitoneal fat deposition (Fig. 10 B-E) significantly increased in HFD group compared to SD counterpart, while in HFD+INF200 group, the deposition of these different types of fat was significantly lower compared with HFD alone. Our data first indicate the effectiveness of HFD regimen, that has been widely shown to induce obesity not only in humans, but also in rodents, where the levels of dietary fats directly correlate with an increase in body weight and body fat deposition [41]. Therefore, both mice and rats are extensively used to model human obesity and metabolic dysfunction, by inducing excessive energy intake and weight gain using hyperlipidic diets, as only increased dietary fat content has been shown to associate with elevated energy intake and adiposity [42, 43], like HFD used in our study that provides ~ 60% total energy intake from fats. Numerous evidence, also deriving from our previous publications, indicates that this hypercaloric regimen triggers a higher degree of obesity affecting glucose metabolism and insulin resistance, and that can be used to generate a valid rat model of obesity, metabolic syndrome with insulin resistance and compromised β-cell function, and metaflammation [44-48]. Although INF200 did not mitigate HFD-induced body weight gain in a significant manner, it should be noted that this NLRP3 inhibitor was administered in the last 4 weeks of HFD [*i.e.,* during the period when Wistar rats usually develop severe obesity [49]; on the other hand, INF200 was able to significantly mitigate HFD-dependent fat deposition, in particular reducing intra-abdominal visceral fat, the deposition of which is combined with unfavorable metabolic activity and an increased risk of cardiovascular complications [50].

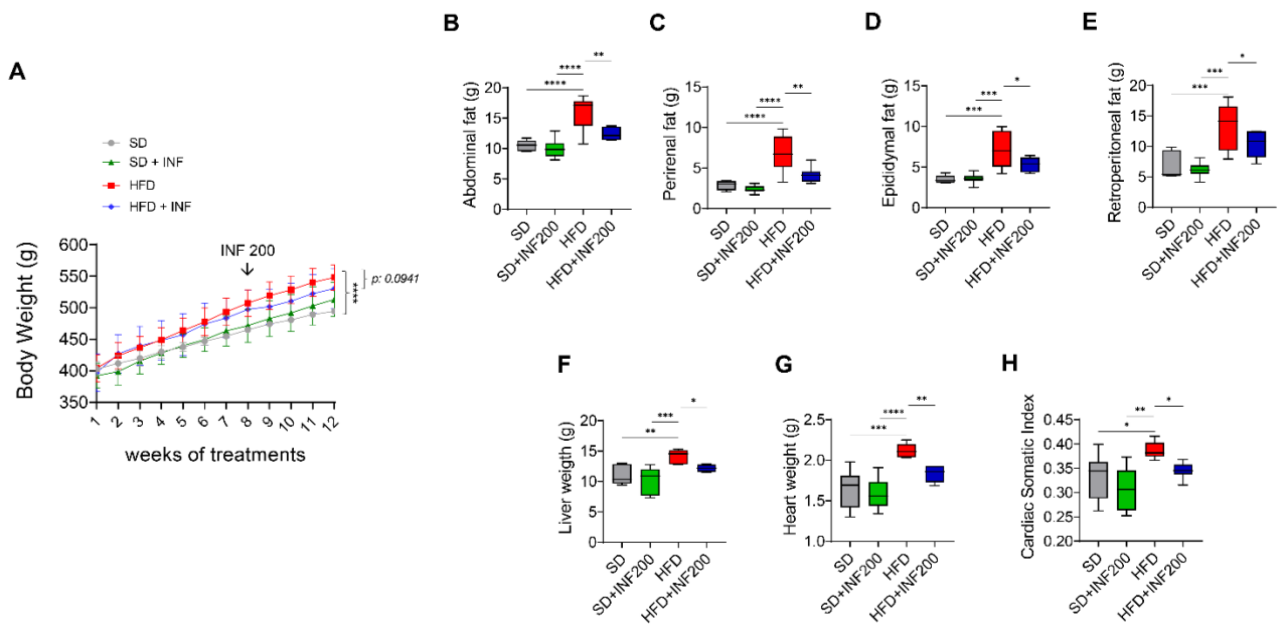

**Figure 10.** Effect of INF200 on «anthropometric» » alterations induced by HFD regimen. A) Body weight in rats fed with SD and treated with vehicle (SD group) (n=6), or with SD and treated with INF200 (SD+INF200 group) (n=7), or with HFD and treated with vehicle (HFD group) (n=6), or with HFD and treated with INF200 (HFD+INF200 group) (n=7) [Bonferroni multiple comparison test, 8,19 % of total variation between groups (p <0,0001)]; B) abdominal; C) perirenal, D) epidydimal, E) retroperitoneal fat, F) liver weight, G) heart weight, H) cardiac somatic index (CSI) in rats fed with SD and treated with vehicle (SD group) (n=6), or with SD and treated with INF200 (SD+INF200 group) (n=7), or with HFD and treated with vehicle (HFD group) (n=6), or with HFD and treated with INF200 (HFD+INF200 group) (n=7); data are expressed as means ± SEM, statistical significance: *£p < 0.05, ** p < 0.01, *** p < 0.001, **** p < 0.0001* (One-way ANOVA and the non-parametric Newman-Keuls Multiple Comparison Test).

The impact of obesity and metabolic syndrome on hepatic and cardiac remodelling and dysfunction is well established. Hepatotoxicity is the main relative factor of hyperlipidaemia and obesity, which are associated with an increased risk of non-alcoholic fatty liver disease (NAFLD), a condition characterized by an increase in intrahepatic triglyceride content with or without inflammation and fibrosis [51]. Diverse studies in animal models and patients reported increased left ventricular mass, ventricular hypertrophy and dysfunction induced by obesity, as result of the increased left ventricular preload and resting cardiac output secondary to the increased total blood volume [52, 53]. In this regard, we found that, in addition to fat weight, also liver weight, heart weight and the relative CSI (i.e. important parameters for attesting the effectiveness of diet-induced obesity) significantly increased during HFD compared to SD. Conversely, all these parameters were significantly reduced in HFD+INF200 group with respect to HFD alone group, without differences between SD and SD+INF200 groups (Fig. 10 F-H), suggesting a possible contribution of INF200 in mitigating obesity-associated pathological hepatic and heart growth.

*2.6.2. INF200 ameliorated HFD-induced metabolic alterations, systemic inflammation, and cardiac dysfunction.*

Metabolic alterations are very common in obesity. Indeed, the altered glucose homeostasis in obese patients depends on a disrupted signaling pathway of insulin resulting in decreased glucose uptake by the muscle, altered lipogenesis, and increased glucose output by the liver [54]. Dyslipidemia is present in almost 70% of obese patients and is usually characterized by high levels of serum free fatty acids, total cholesterol, triglycerides, very low-density lipoproteins (VLDL), and low levels of HDL-cholesterol together with HDL dysfunction [55]. The saturated fats present in HFD regimens used to induce experimental obesity are also responsible for the alteration in glucose and lipid profiles. According to this evidence, we investigated the metabolic impact of INF200 during HFD. Our analyses revealed a significant increase of glycaemia, plasma levels of total cholesterol, LDL cholesterol, and triglycerides together with low plasma levels of HDL cholesterol in the HFD group compared to the SD group (Fig. 11 A-E). On the contrary, the treatment with INF200 significantly ameliorated the glyco-metabolic profile, relieving the HFD-dependent alteration in all the above indexes. In the SD+INF200 group, total cholesterol, LDL cholesterol, triglycerides, and HDL cholesterol did not significantly change compared with the SD control group (Fig. 11 A-E).

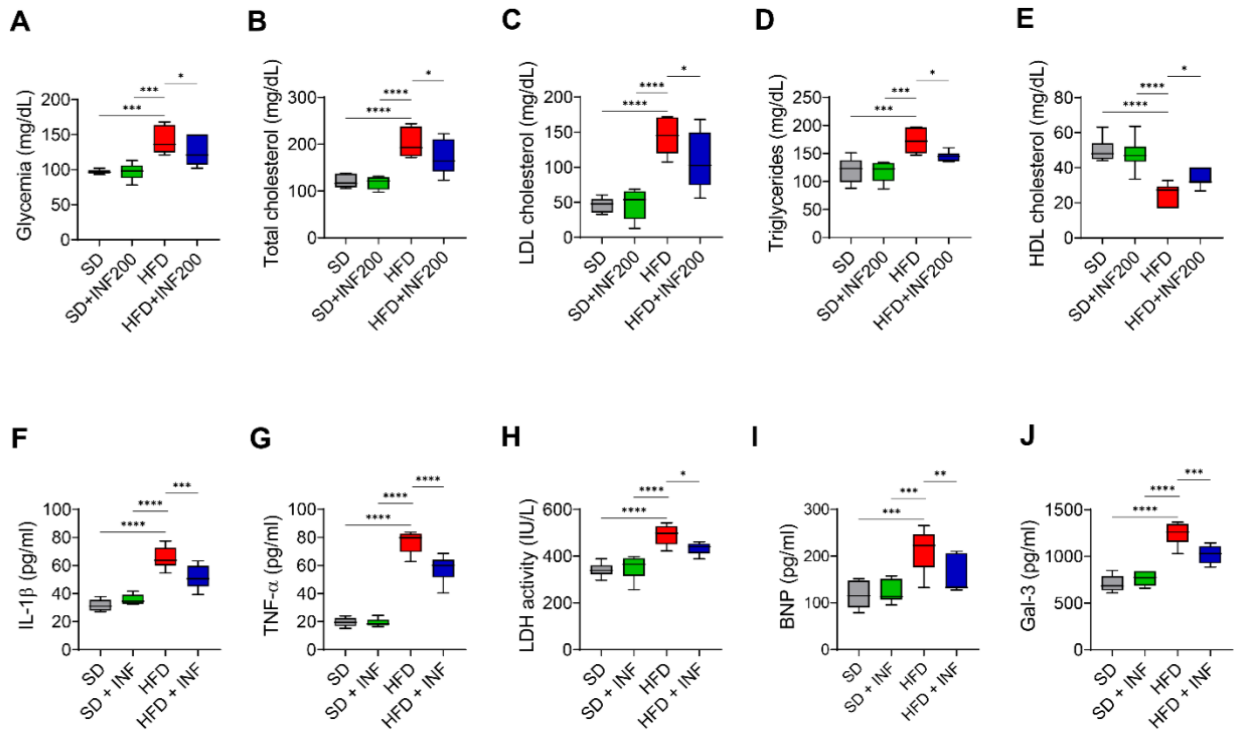

**Figure 11.** Influence of INF200 on HFD-induced metabolic alterations, systemic inflammation, and cardiac dysfunction. Plasma levels of A) glucose, B) total cholesterol, C) LDL cholesterol, D) triglycerides, E) HDL cholesterol, F) IL-1β, G) TNF-α, H) LDH, I) BNP, J) Gal-3 in rats fed with SD and treated with vehicle (SD group) (n=6), or with SD and treated with INF200 (SD+INF200 group) (n=7), or with HFD and treated with vehicle (HFD group) (n=6), or with HFD and treated with INF200 (HFD+INF200 group); data are expressed as means ± SEM statistical significance: * $p < 0.05$, ** $p < 0.01$, *** $p < 0.001$, **** $p < 0.0001$ (One-way ANOVA and the non-parametric Newman-Keuls Multiple Comparison Test).

All the obesity-related glucose and lipid abnormalities, together with the increase in adipose tissue mass that alters adipokine secretion pattern, are frequent in metabolic syndrome and typically associate with a pro-inflammatory state and metaflammation, which predispose to CVD [56]. Indeed, the excess of nutrients inducing metabolic inflammation is not confined to adipose tissue, but the obesity-related influx of immune cells occurs in many other tissues that produce inflammatory mediators (e.g., cytokines and chemotactic molecules) that influence immune cells' action and compromise the physiology of several organs including the heart [14, 57]. It is now clear that elevated systemic levels of pro-inflammatory cytokines and metabolic substrates induce an inflammatory state in different cardiac cells and lead to subcellular alterations thereby promoting maladaptive myocardial remodelling and dysfunction [58]. To assess the effect of INF200 on HFD-dependent systemic inflammation and myocardial damage, we first monitored the serum release of the key pro-inflammatory cytokines IL-1β and TNF-α, and then analysed the serum levels of specific biomarkers of heart failure, such as LDH, BNP and Gal-3. Our results indicated a significant increase of all these markers in rats fed with HFD compared to SD group and a significant action of INF200 in mitigating their levels in HFD conditions (Fig. 11 F-J). No significant differences have been detected among SD and SD+INF200 groups (Fig. 11 F-J). Accordingly, during obesity the altered lipid profile increases pro-inflammatory cytokines, such as IL-6, IL-1β and TNF-α, whose levels correlate with the rise of BMI [59]. Several inflammatory diseases correlated with serum LDH and emerging studies also supported a positive association between higher level of serum LDH and mortality from all causes in individuals with metabolic syndrome [60]. On the other hand, Gal-3, a member of a β-galactoside–binding lectin family, represents a recognized mediator of cardiac damage through stimulation of ECM deposition and exacerbating inflammatory state [61, 62]. Interestingly, high levels of myocardial and serum Gal-3 levels have also been reported in both diet-induced obesity rat models and obese patients, playing a key role in cardiovascular remodeling associated with obesity [63-65]. In addition, clinical evidence indicates that Gal-3 levels markedly associate with outcomes in HF patients with preserved ejection fraction (HFpEF) compared with HF patients with reduced ejection fraction (HFrEF) [66], and increase over time in more severe HF and renal dysfunction patients associating with a poorer clinical outcome [67]. Noteworthy, the combination of Gal-3 with the aminoterminal portion of pro–brain natriuretic peptide (NT-proBNP) represents the best predictor for prognosis in individuals with acute HF [68, 69]. BNP is mainly generated by the heart in response to myocardial stress and represents a specific indicator of heart disease severity. Therefore, our data suggest that INF200 can relieve inflammatory response triggered by metabolic disturbance during obesity that impairs myocardial function leading to metabolic cardiomyopathy.

*2.6.3. INF200 limited IRI dependent contractile impairment and infarct size in both normal and obese conditions.*

Growing evidence indicates that obesity and insulin-resistance not only represent major risk factor for the development of CVD, including ischemic heart disease, but they also induce adverse cardiac remodelling and LV pump dysfunction associated with increased susceptibility to myocardial IRI [70-72]. Therefore, we tested whether INF200 could affect the outcome of myocardial IRI in *ex vivo* settings in both normal (SD) and obese (HFD) conditions. In particular, we assessed the potential cardioprotective action of INF200 against IRI by first evaluating the post-ischemic systolic and diastolic recovery and then assessing the extent of myocardial infarction through evaluation of ischemia-induced LDH release in coronary effluents and myocardial infarct size (IS).

Figure 12A, indicating the values of the developed LVP recovery (dLVP; i.e., the inotropic activity) before ischemia, during reperfusion and at the end of reperfusion, shows that dLVP was significantly lower than before ischemia in both SD and HFD groups (*SD group: dLVP: 43 ± 7 mmHg at the end of reperfusion; baseline values: 69 ± 4 mmHg; HFD group: dLVP: 32 ± 4 mmHg at the end of reperfusion; baseline values: 54 ± 6 mmHg*). In contrast, in SD+INF200 and HFD+INF200 groups, dLVP recovery significantly improved during reperfusion and at the end of reperfusion (*SD+INF200 group: dLVP: 80 ± 5 mmHg at the end of reperfusion; baseline values: 65 ± 5 mmHg; HFD+INF200 group: dLVP: 65 ± 10 mmHg at the end of reperfusion; baseline values: 64 ± 5 mmHg*), compared with their control counterparts (SD alone and HFD alone, respectively).

We then evaluated left ventricular end-diastolic pressure (LVEDP), as an important index of diastolic function before ischemia and in the post-ischemic phase, where LVEDP of 4 mmHg or more above baseline level represents an important indicator of cardiac damage in the rat heart [73-75]. Our results showed a significant increase of LVEDP values in SD and HFD groups during the reperfusion (*SD group: LVEDP: 21 ± 2 mmHg at the end of reperfusion; baseline values: 3 ± 1 mmHg; HFD group: LVEDP: 27 ± 1 mmHg at the end of reperfusion; baseline values: 6 ± 0.2 mmHg*). On the contrary, in the SD + INF200 and HFD + INF200 groups, LVEDP was significantly lower compared to SD and HFD alone, respectively (*SD + INF200 group: LVEDP: 13 ± 0.7 mmHg at the end of reperfusion; baseline values: 3 ± 0.9 mmHg; HFD + INF200 group: LVEDP: 18 ± 2 mmHg at the end of reperfusion; baseline values: 4 ± 0.6 mmHg*). Notably, in HFD hearts, cardiac contracture was more evident, as revealed by a significant elevation of LVEDP at the end of reperfusion in HFD group with respect to SD group (Fig. 12B).

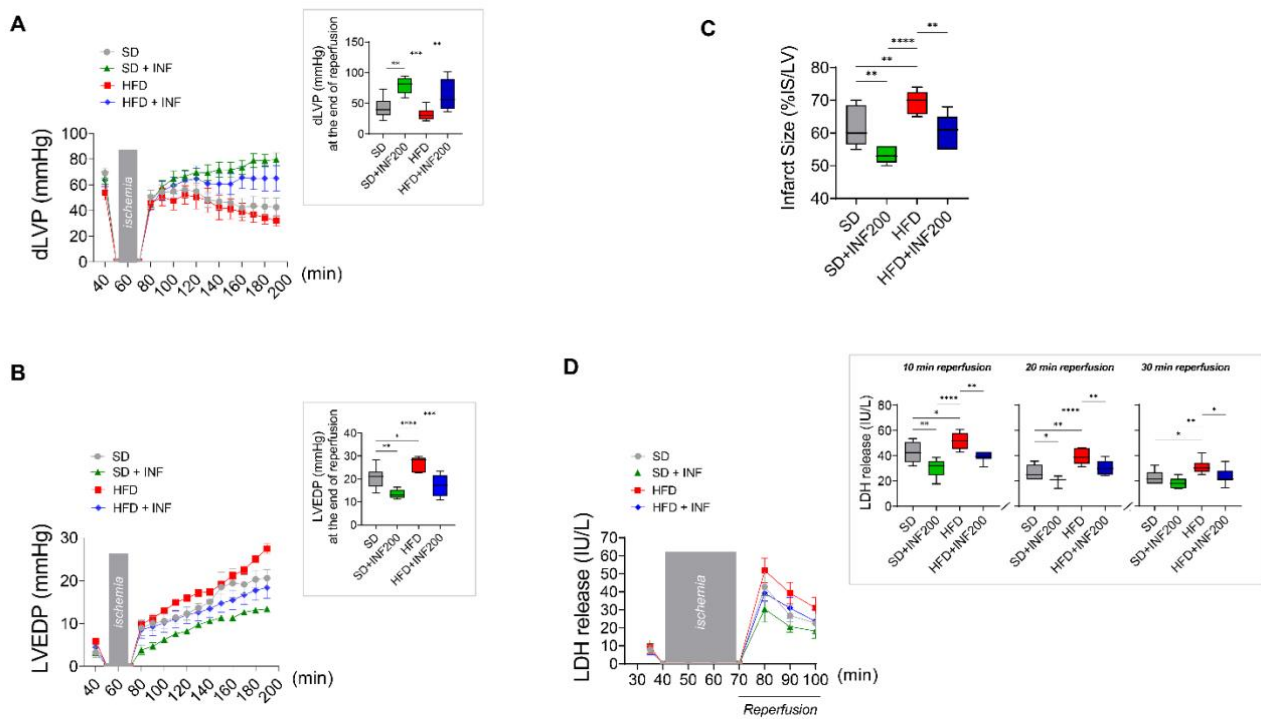

**Figure 12.** Cardioprotective potential of INF200. Effects of INF200 on systolic and diastolic function of Langendorff perfused normal and obese hearts subjected to I/R injury (IRI), on infarct size and release of LDH in coronary effluent. A) dLVP and B) LVEDP variations. Gray boxes indicate ischemic administration (Bonferroni multiple comparison test). dLVP = 7.45 % of total variation between groups (p <0.0001), LVEDP = 9.17 % of total variation between groups (p <0.0001). Inset graph shows dLVP and LVEDP at the end of reperfusion. Data are expressed as changes of dLVP and LVEDP values (millimeters of mercury) from stabilization to the end of the 120 min of reperfusion with respect to the baseline values for rats fed with SD and treated with vehicle (SD group) (n=6), or with SD and treated with INF200 (SD+INF200 group) (n=7), or with HFD and treated with vehicle (HFD group) (n=6), or with HFD and treated with INF200 (HFD+INF200 group) (n=7); * $p < 0.05$, ** $p < 0.01$, *** $p < 0.001$, **** $p < 0.0001$ (One-way ANOVA and the non-parametric Newman-Keuls Multiple Comparison Test). C) Infarct size (IS). The amount of necrotic tissue measured after 30 min global ischemia and 120 min reperfusion is expressed as a percentage of the LV mass (% IS/LV) in IRI hearts from rats fed with SD (n=6), or SD+INF200 (n=7), or HFD (n=6), or HFD+INF200 (n=7); ** $p < 0.01$, **** $p < 0.0001$ (1-way ANOVA and Newman-Keuls multiple comparison test). D) LDH activity was assessed in coronary effluent 5 min before ischemia and 10, 20 and 30 min after ischemia in the reperfusion phase. LDH variations in IRI hearts from rats fed with SD (n=6), or SD+INF200 (n=7), or HFD (n=6), or HFD+INF200 (n=7) (Bonferroni multiple comparison test, 2.13 % of total variation between groups (p <0.0001). Inset graph shows LDH activity at the end of 10, 20 and 30 min of reperfusion. Data are expressed as IU/L, * *p < 0.05*, ** *p < 0.01*, **** *p < 0.0001* (one-way ANOVA and Newman–Keuls multiple comparison test).

An analogous trend was observed in assessing the extent of IS (expressed as a percentage of LV mass) and LDH release in coronary effluents. In fact, as depicted in Fig. 12C, IS was significantly higher in the groups not treated with INF200, with a further increase observed in the hearts of HFD rats compared with those of SD rats *(~ 62 ± 2 % of IS/LV in SD group and ~ 70 ± 1 % of IS/LV in HFD group)*. Conversely, a significant reduction of IS has been found in both SD and HFD hearts treated with INF200 compared to their control counterparts SD and HFD alone, respectively *(~ 53 ± 1 % of IS/LV in SD+INF200 group and ~ 61 ± 2 % of IS/LV in HFD+INF200 group)* (Fig. 12C). Notably, the measurement of IS represents a key element for establishing the clinical outcomes in patients with post-ischemic HF, as well as an attractive surrogate end point for the early assessment of new therapies for acute myocardial infarction [76]. Therefore, selective strategies able to decrease the extent of IS show clinical interest.

The activity of LDH in coronary effluents (collected prior to ischemia and during the first 10, 20, and 30 min of reperfusion [77, 78]) was assessed to provide further indication on the extent of myocardial cellular damage during reperfusion. As showed in Fig. 12D, that indicates LDH release during the first 30 min of reperfusion and at 10, 20, and 30 min of reperfusion, the enzyme activity was significantly higher in SD and HFD groups at all times of reperfusion phase. Also in this case, in HFD hearts, LDH release was more pronounced than that of SD hearts at 10 min and 20 min of reperfusion. Differently, INF200 treatment relieved LDH levels during all the 30 min of reperfusion, both in SD (SD+INF200 group) and HFD (HFD+INF200) conditions with respect to SD and HFD alone groups, indicating a protective role of INF200 against necrotic damage with a similar pattern to the recovery of haemodynamic function.

Overall, these results first indicate that, following myocardial IRI, the obese rats showed exacerbation of reperfusion injury and a more severe impairment of myocardial function compared with the non-obese rats, highlighting that obesity can aggravate the susceptibility to myocardial infarction and myocardial IRI. On the other hand, our data suggest that improving metabolic and inflammatory dysfunction by pharmacological intervention, aimed at inhibiting NLRP3, could ameliorate reperfusion-induced myocardial injury.

*2.4.4. INF200 mitigated IRI-provoked NLRP3 activation, inflammation and oxidative stress in normal and obese hearts.*

Although the molecular mechanisms responsible for the exacerbation of myocardial IRI in obese and insulin-resistance conditions have not been fully elucidated, diverse pre-clinical and clinical data indicate that excessive ROS generation, the impairment of anti-oxidant endogenous capacities, alteration in insulin signaling and increased inflammation, play an important role. All these events, together with metaflammation occurring during the metabolic cardiomyopathy lead to impaired myocardial relaxation, and in the advanced stage, the vicious cycle of subcellular component abnormalities results in impairment of both diastolic and systolic functions [79]. In this regard, we extended our study to determine the potential ability of INF200 to inhibit IRI-dependent activation of NLRP3 activation, inflammation and oxidative stress in normal and obese hearts. To this aim, we carried out western blot and biochemical analyses on LV extracts after IRI protocols. Immunoblot and relative densitometric analysis showed a significant increase of NLRP3 protein expression in both SD and HFD groups subjected to IRI compared to the sham hearts. In contrast, a significant reduction in the NLRP3 expression levels was observed in SD+INF200 and HFD+INF200 groups with respect to their relative controls (SD and HFD alone, respectively) (Fig. 13A).

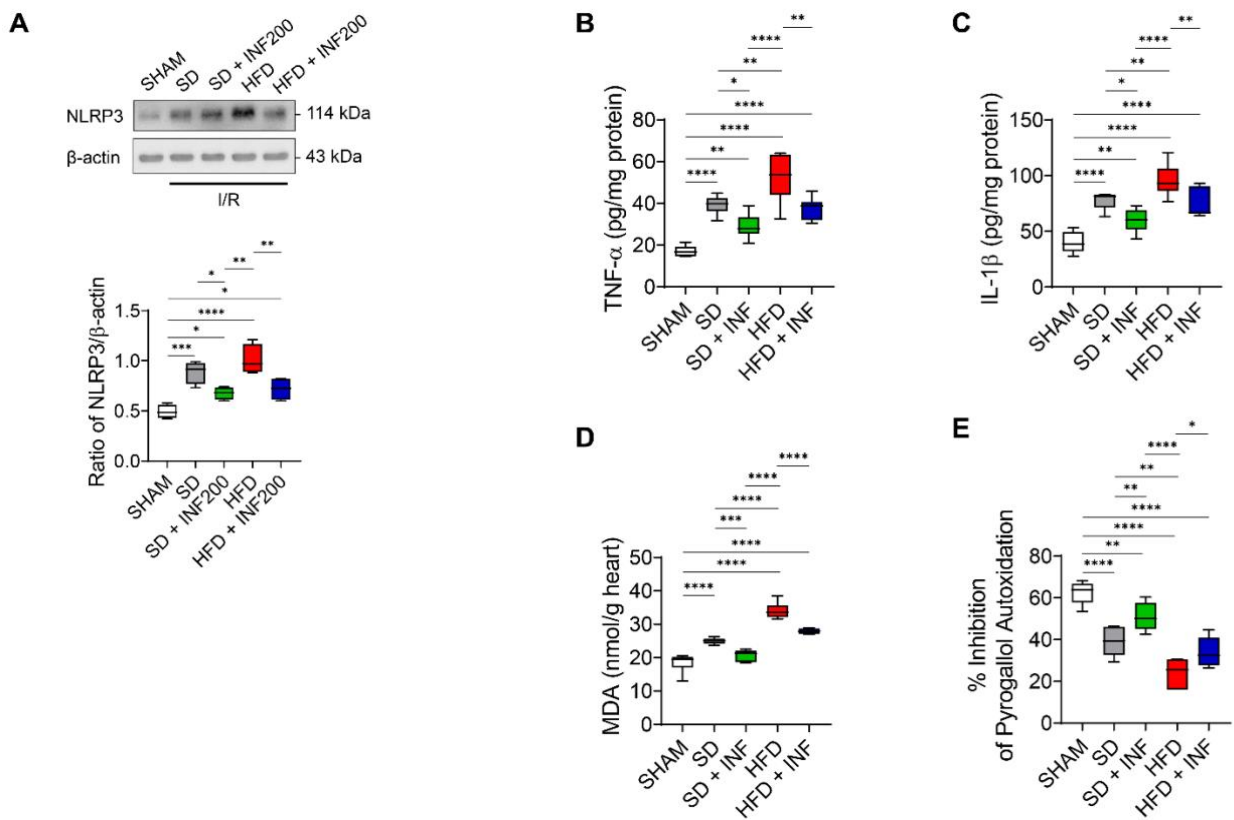

**Figure 13.** Effect of INF200 on IRI-dependent cardiac activation of NLRP3, inflammation and oxidative stress in normal and obese conditions. A) Western Blot of NLRP3 expression in cardiac extracts of sham and rats fed with SD and treated with vehicle (SD group), or with SD and treated with INF200 (SD+INF200 group), or with HFD and treated with vehicle (HFD group), or with HFD and treated with INF200 (HFD+INF200 group) exposed to IRI protocols (n=4 for each group). Graph represent the ratio of densitometric analysis of NLRP3/β-actin; data are expressed as means ± SEM; statistical significance: *$p < 0.05$, **$p < 0,01$, ***$p < 0,001$, ****$p < 0.0001$ (One-way ANOVA and the non-parametric Newman-Keuls Multiple Comparison Test). Cardiac levels of B) TNF-α, C) IL-1β, D) MDA and E) SOD expressed as % of pyrogallol autoxidation of sham hearts (n=6), and IRI hearts from rats fed with SD (n=6), or SD+INF200 (n=7), or HFD (n=6), or HFD+INF200 (n=7); data are expressed as means ± SEM; statistical significance: *$p < 0.05$, **$p < 0,01$, ***$p < 0,001$, ****$p < 0.0001$ (One-way ANOVA and the non-parametric Newman-Keuls Multiple Comparison Test).

On the other hand, Fig. 13B-C shows a significant increase of myocardial production of IL-1β and TNF-α in SD and HFD groups compared to the sham hearts, whereas the treatment with INF200 was able to significantly decrease these cardiac pro-inflammatory markers in both SD and HFD groups. Additionally, to evaluate the potential effect of INF200 in reducing cardiac oxidative stress, we evaluated the cardiac levels of MDA (i.e. a marker of lipid peroxidation following ROS generation) and the activity of SOD (a crucial endogenous antioxidant enzyme). Results showed a significant increase of MDA levels in SD and HFD groups than those found in sham group, whereas the treatment with INF200 significantly attenuated cardiac MDA generation in both SD and HFD groups (Fig. 13D). Conversely, SOD enzymatic activity significantly decreased in SD and HFD groups with respect to sham hearts, while it was significantly higher in the SD+INF200 and HFD+INF200 groups compared to SD and HFD control counterparts (Fig. 13E). Accordingly, the contribution of ROS and oxidative stress to myocardial IRI is well established. During IRI, especially reperfusion, excessive ROS levels generated from different sources, including increased xanthine oxidase, neutrophil respiratory burst and altered mitochondrial electron transport chain, can induce macromolecule damage, including lipid peroxidation and affect the endogenous antioxidant defence, reflected by decreased activity of crucial antioxidant enzymes, such as SOD [80-82]. On the whole, these data indicate that systemic inhibition of NLRP3 may also reduce IRI-induced intracardiac upregulation of NLRP3, as a major mediator of inflammatory responses associated with several toxic signals, including ROS, which induce oxidative stress and an inflammatory cascade driven by proinflammatory cytokines [83]. In particular, NLRP3 inflammasome is responsible for the cleavage of procaspase-1 to active caspase-1, which in turn induces processing and maturation of IL-1β, an early mediator of inflammasome and pyroptosis- processes driving myocardial IRI [84].

Therefore, the amelioration of cardio-specific inflammation induced by NLRP3 activation and oxidative stress following INF200 treatment, further describes this inhibitor as a promising candidate in the treatment of IRI and its exacerbation during obesity and metabolic cardiomyopathy.

## 3. Conclusions

In this work we designed new non-sulfonylurea-based NLRP3 inhibitors by replacing the sulfonylurea moiety with different heterocycles. Computational studies allowed the design of three series of compounds potentially able to bind the NACHT domain of NLRP3. *In vitro* studies in human macrophages allowed the identification of the 1,3,4-oxadiazol-2-one derivative **5** (INF200) as the most promising NLRP3 inhibitor of the series. INF200 was tested in a model of high-fat diet-induced metaflammation to investigate its effect on cardiometabolic functions. Our data confirm a close relationship between NLRP3, metabolism, and cardiometabolic disorders causing inflammation in the context of obesity and metabolic dysfunction and highlight an important role of this inflammasome in IRI. The use of the inhibitor we synthesized allowed us to understand and consolidate some of the molecular and inflammatory mechanisms of myocardial and systemic changes. The novel 1,3,4-oxadiazole-2-one NLRP3 inhibitor, INF200, which has been now validated for its efficacy in reducing systemic and myocardial changes *in vitro* and *in vivo* in a rat model of HFD-induced metaflammation. These new data pave the way for further improvements in the field of these inhibitors for future application in the clinical field.

## 4. Materials and methods

*4.1. General experimental procedures*

All the reactions were monitored by Thin Layer chromatography (TLC) on Merck 60 F254 (0.25 mm) plates, which were visualised by UV inspection (254 nm) and/or by spraying KMnO$_4$ (0.5 g in 100 mL 0.1 N NaOH). Na$_2$SO$_4$ was used as drying agent for the organic phases. Flash chromatography (FC) purifications were performed using silica gel Merck with 60 mesh particles. Unless otherwise specified, all reagents were used as received without further purification. Dichloromethane was dried over P$_2$O$_5$ and freshly distilled under nitrogen prior to use. DMF was stored over 3 Å molecular sieves. Anhydrous THF was freshly distilled under nitrogen from Na/benzophenone ketyl. $^1$H and $^{13}$C-NMR spectra were registered on JEOL ECZR600 spectrometer, at 600 and 151 MHz. Coupling constants (J) are given in Hertz (Hz) and chemical shifts (δ) are given in ppm, calibrated to solvent signal as internal standard. Following abbreviations are used to describe multiplicities: s= singlet, d = doublet, t = triplet, q = quadruplet, m = multiplet and br= broad signal. The following abbreviations are used to identify exact proton: ArH = Aromatic proton, BzImH= benzimidazolone ring, Pip = piperidine. ESI-mass spectra were recorded on a Waters Micromass Quattro Micro equipped with an ESI source. The purity of the final compounds was determined by RP-HPLC. Analyses were performed with a HP1100 chromatograph system (Agilent Technologies, Palo Alto, CA, USA) equipped with a quaternary pump (G1311A), a membrane degasser (G1379A) and a diode-array detector (DAD) (G1315B) integrated into the HP1100 system. Data analyses were processed using a HP ChemStation system (Agilent Technologies). The analytical column was a LiChrospher 100 C18-e (250 × 4.6 mm, 5 μm) (Merck KGaA, 64271 Darmstadt, Germany) eluted with CH$_3$CN 0.1% TFA/H$_2$O 0.1% TFA in a ratio that depended on the characteristics of the compound. All compounds were dissolved in the mobile phase at a concentration of about 0.01 mg/mL and injected through a 20 μL loop. HPLC retention times (tR) were obtained at flow rates of 1.0 mL/min and the column effluent was monitored by DAD at 226 and 254 nm (with 800 nm as the reference wavelength). The purity of the samples was evaluated as the percentage ratio between the areas of the main peak and of possible impurities at the two wavelengths, and also using a DAD purity analysis of the chromatographic peak. The purity of all the target compounds was found to be greater than 95% (Table S1). NP3-146 was synthesized according to the reported procedure [29]

*4.2. Synthesis of compounds*

**N'-Hydroxy-2-(3-(trifluoromethyl)phenyl)acetimidamide) (15)** to a solution of 3-trifluoromethyl phenyl acetonitrile (1 g, 5.40 mmol ) in EtOH 96% (15 mL), hydroxylamine hydrochloride (777 mg, 11.88 mmol) and TEA (1.73 mL, 12.42 mmol) were added and the mixture was heated at 75 °C. After 18 h the solvent was evaporated under reduced pressure and the residue was treated with NaOH 1 M. The aqueous phase was extracted with ethyl acetate (6 × 50 mL) and the combined organic phases were washed with brine (30 mL), dried (Na$_2$SO$_4$) and concentrated under reduced pressure to give **15** as white solid (1.03 g, 87.5%). Rf = 0.15 (petroleum ether/ethyl acetate 8:2). MS (ESI$^+$): *m/z* 219 [M+H]$^+$; $^1$H-NMR (600 MHz, CD$_3$CN) δ 7.72 (s, 1H), 7.65 (d, J = 7.6 Hz, 1H), 7.59 (d, J = 7.9 Hz, 1H), 7.50 (t, J = 7.7 Hz, 1H), 3.81 (s, 2H), 3.08 (s, 1H). $^{13}$C

NMR (151 MHz, CDCl$_3$). 13C-NMR (151 MHz, CD$_3$CN) δ 159.7, 135.3, 133.0, 130.4 (q, J = 31.8 Hz), 129.8, 125.9, 124.6, 124.3 (q, J = 271.2 Hz), 44.1.

**Methyl 4-(3-(3-(trifluoromethyl)benzyl)-1,2,4-oxadiazol-5-yl)benzoate (1)** Compound **15** (500 mg, 2.28 mmol) was dissolved in toluene (50 mL) then methyl 4-fromyl benzoate (930 mg, 2.51 mmol) and PTSA (45 mg, 0.229 mmol) were added and heated under reflux for 24 h using a Dean-Stark trap to remove the water. The reaction mixture was stirred at room temperature for 3 h. The solvent was evaporated under reduced pression, the obtained residue treated with water (50 mL) and extracted with DCM (3 × 50 mL). The combined organic phases were washed with brine (150 mL), dried (Na$_2$SO$_4$) and concentrated under reduced pressure. The crude product was purified by silica gel chromatography (petroleum ether /ethyl acetate 95:5) and then recrystallized from ethanol to give **1** (130 mg, 15.2%) as a white solid. Rf=0.23 (petroleum ether/ethyl acetate 9:1). MS(ESI$^+$) *m/z*: 363 [M+H]$^+$. $^1$H NMR (600 MHz, CDCl$_3$) δ=8.17 (d, J = 1.7 Hz, 4H, ArH), 7.64 (s, 1H, CF$_3$ArH$^2$), 7.57 (d, J = 7.6 Hz, 1H, CF$_3$ArH$^4$), 7.53 (d, J = 7.8 Hz, 1H, CF$_3$ArH$^6$), 7.46 (t, J = 7.7 Hz, 1H, CF$_3$ArH$^5$), 4.20 (s, 2H, CH$_2$), 3.95 (s, 3H, CH$_3$). $^{13}$C NMR (151 MHz, CDCl$_3$) δ=175.25, 169.77, 166.11, 136.36, 133.93, 132.59, 131.21 (q, J = 32.1 Hz, CF$_3$*C*), 130.37, 129.36, 128.25, 127.78, 126.01 (d, J = 3.9 Hz, CF$_3$C*C*H), 124.31 (CF$_3$C4'CH), 124.11 (q, J = 272 Hz, CF$_3$), 52.70, 32.34.

**4-(3-(3-(Trifluoromethyl)benzyl)-1,2,4-oxadiazol-5-yl)benzoic acid (2).** To a solution of **15** (10 mg, 0.028 mmol) in THF (0.3 mL), LiOH 1M (1.38 μL, 0.138 mmol) were added and the obtained mixture was kept under stirring at rt. After 18 h the mixture was treated with HCl 1M (100 μL) to precipitate the reaction produict as a white solid. The solid was treated with water (100 μL) and collected by vacuum filtration. The obtained solid was washed with cool water to give **2** as a white solid (5 mg, 63.1%). MS (ESI$^-$): *m/z* 347 [M-H]$^-$; $^1$H-NMR (600 MHz, DMSO-D$_6$) δ= 13.50 (s, br, 1H, COOH), 8.16-8.14 (m, 2H, Ar-H), 8.10-8.08 (m, 2H, Ar-H), 7.74 (s, 1H, Ar-H), 7.66-7.55 (m, 3H, Ar-H), 4.32 (s, 2H, CH$_2$). $^{13}$C-NMR (151 MHz, DMSO-D$_6$) δ= 175.04, 170.30, 166.91, 137.61, 135.23, 133.84, 130.78, 130.21, 129.78 (q, J= 31.2 Hz),128.64, 127.25, 126.22, 124.68 (q, J= 271 Hz), 31.53.

**Methyl-4-(*N'*-hydroxycarbamimidoyl) benzoate (17).** To a stirred solution of methyl-4-cianobenzoate (500 mg, 3.10 mmol) in EtOH 96% (7ml), hydroxylamine hydrochloride (473 mg, 6.82 mmol) and TEA (0.995 mL,7.13mmol) were added and the mixture was heated under reflux for 3 h. The solvent was evaporated under reduced pression and the obtained residue purified by silica gel chromatography (DCM/MEOH 98:2 to 95:5) to give **17** (530 mg, 87.2%) as a white solid. Rf=0.22 (DCM/MeOH 98:2). $^1$H NMR (600 MHz, CDCl$_3$) δ= 8.01 (dd, J = 8.5, 1.9 Hz, 2H, ArH$^{3,'5'}$), 7.65 (dd, J = 8.5, 1.9 Hz, 2H, ArH$^{2,'6'}$), 3.88 (s, 3H, OCH$_3$), 2.73 (br, 2H, NH$_2$).

**Methyl 4-(5-(2-chlorobenzyl)-1,2,4-oxadiazol-3-yl)benzoate (3).** To a stirred solution of 2-chlorophenylacetic acid (240 mg, 1.41 mmol) in dry THF (3,0 mL), CDI (228 mg, 1.41 mmol) was added and the mixture was stirred at rt for 1 h before addition of **17** (250 mg, 1.29 mmol). The reaction mixture was stirred at rt for 3 h. Evaporation of solvent under reduced pressure afforded a white solid which was dissolved in acetic acid and heated under reflux 18 h. Once the mixture reached rt, Na$_2$CO$_3$ 10% solution (20 mL) was added, and the product was extracted with DCM (3 × 20 mL). The combined organic phases were washed with brine (30 mL), dried (Na$_2$SO$_4$) and concentrated under reduced pressure. The crude product crystallized from ethanol to give **3** (260 mg, 61%) as a white solid. Rf=0.21 (petroleum ether/ethyl acetate 9:1). MS (ESI$^+$): *m/z* 329 / 331 [M+H]$^+$; $^1$H-NMR (600 MHz, CDCl$_3$) δ 8.14-8.11 (m, 4H), 7.42 (d, J = 4.1 Hz, 1H), 7.37 (t, J = 4.6 Hz, 1H), 7.28-7.25 (m, 2H), 4.43 (s, 2H), 3.93 (s, 3H). $^{13}$C-NMR (151 MHz, CDCl$_3$) δ 177.6, 168.0, 166.5, 134.4, 132.5, 131.6, 131.1, 130.9, 130.1, 130.0, 129.4, 127.6, 127.4, 52.5, 31.1.

**Methyl 4-(5-(3-(trifluoromethyl)benzyl)-1,2,4-oxadiazol-3-yl)benzoate (4).** To a stirred solution of 3-trifluoromethylphenylacetic acid (289 mg, 1.41 mmol) in dry THF (3,0 mL), CDI (228 mg, 1.41 mmol) was added and the mixture was stirred at rt for 1 h before addition of **17** (250 mg, 1.29 mmol). The reaction mixture was stirred at rt for 3 h. Evaporation of solvent under reduced pression afforded a white solid which was dissolved in acetic acid and heated under reflux 18 h. Once the mixture reached rt, Na$_2$CO$_3$ 10% solution (20 mL) was added, and the product was extracted with DCM (3 × 20 mL). The combined organic phases were washed with brine (30 mL), dried (Na$_2$SO$_4$) and concentrated under reduced pressure. The crude product crystallized from ethanol to give **4** (255 mg, 55%) as a white solid. Rf=0.21 (petroleum ether/ethyl acetate 9:1). MS (ESI$^+$): *m/z* 363 [M+H]$^+$; $^1$H NMR (600 MHz, CDCl$_3$) δ=8.19 – 8.09 (m, 4H, ArH), 7.59 (m, 4H, CF$_3$ArH), 4.37 (s, 2H, CH$_2$), 3.95 (s, 3H, OCH$_3$). $^{13}$C NMR (151 MHz, CDCl$_3$) δ 177.6, 168.0, 166.5, 134.2, 132.6, 132.5, 131.5 (q, J = 32.3 Hz), 130.7, 130.2, 129.6, 127.5, 126.0, 124.8, 124.1 (q, J = 272.1 Hz), 52.5, 32.9.

**2-(2-Chlorophenyl) acetohydrazide (19).** To a stirred solution of 2-chlorophenylacetic acid (**18**, 1 g, 5.86 mmol) in THF (15 mL), CDI (1.046 g, 6.45 mmol). After 1 hour at rt, hydrazine monohydrate (0.440 g, 8.79 mmol) was added and the reaction was left under stirring overnight. The crude product was purified by silica gel chromatography (DCM/MeOH 95:5) to give **19** (0.920 g, 85.2%) as a white opalescent solid. Rf= 0.18 (DCM/MeOH from 98:2 to 95:5). MS (ESI$^+$): *m/z* 185 / 187 [M+H]$^+$; $^1$H NMR (600 MHz, CDCl$_3$) δ=7.37 – 7.33 (m, 1H, ArH$^6$), 7.31 – 7.28 (m, 1H, ArH$^3$), 7.24 – 7.19 (m, 2H, ArH$^{4,5}$), 3.63 (s, 2H, CH$_2$), 2.66 – 2.51 (m, 2H, NH$_2$). $^{13}$C NMR (151 MHz, CDCl$_3$) δ= 170.8, 134.4, 132.2, 131.7, 129.8, 129.1, 127.4, 39.3.

**5-(2-Chlorobenzyl)-1,3,4-oxadiazol-2(3*H*)-one (20).** The hydrazide **19** (0.630 g, 3.42 mmol) and CDI (0.610 g, 3.76 mmol) were dissolved in dry THF (60 mL) and stirred at rt ovenight under nitrogen atmosphere. The solvent was evaporated and the obtained residue taken up with a water (25 mL) and extracted with ethyl acetate (3 × 20 mL). The combined organic phases were washed with brine (25 mL), dried (Na$_2$SO$_4$) and concentrated under reduced pressure. The crude product was purified by silica gel chromatography (DCM/MEOH 99:1 to 95:5) to give **20** (648 mg, 90.3%) as a white solid. Rf=0.20 (DCM/MeOH 98:2). MS (ESI$^+$): *m/z* 211 / 213 [M+H]$^+$; $^1$H NMR (600 MHz, CDCl$_3$) δ= 9.76 (s, 1H, NH), 7.44 –7.36 (m, 1H, ArH$^6$), δ=7.33 – 7.27 (m, 1H, ArH$^3$), 7.27 – 7.21 (m, 2H, ArH$^{4,5}$), 4.02 (s, 2H, CH$_2$). $^{13}$C NMR (151 MHz, CDCl$_3$) δ=155.85, 155.61, 134.39, 131.16, 130.74, 130.01, 129.54, 127.40, 30.78.

**General procedure for the synthesis of compounds 5, 7, 9, and 21.** To a stirred solution of intermediate **20** in dry THF kept at 0 °C, DBU (1.5 mmol) was added and stirring was continued for 30 minutes. After this time, the desired bromo derivative was added to the reaction mixture obtaining the formation of a white precipitate. The reaction mixture was stirred at rt for 4 h. The liquid phase was transferred in another flask and concentrated under reduced pressure. The residue was treated with water (10 mL) and extracted with ethyl acetate (3 × 10 mL). The combined organic phases were washed with brine (30 mL), dried ($Na_2SO_4$) and concentrated under reduced pressure. The crude product was purified as described below.

**Ethyl 2-(5-(2-chlorobenzyl)-2-oxo-1,3,4-oxadiazol-3(2*H*)-yl)acetate (5).** The reaction was run with compound **20** (100 mg, 0.475 mmol), DBU (142 μL, 0.950 mmol) and ethyl 2-bromoacetate (78 μL, 0.712 mmol) in dry THF (2 mL). The crude product was purified by silica gel chromatography (petroleum ether/ethyl acetate 95:5 to 85:15) to give **5** (100 mg, 70.9%) as a yellowish oil. Rf = 0.25 (petroleum ether/ethyl acetate 8:2). MS (ESI$^+$): *m/z* 297 / 299 [M+H]$^+$; $^1$H NMR (600 MHz, CDCl$_3$) δ=7.43 – 7.40 (m, 1H, ArH$^6$), 7.32 (m, 1H, ArH$^3$), 7.29 – 7.26 (m, 2H, ArH$^{4,5}$), 4.44 (s, 2H, NCH$_2$), 4.24 (t, J = 7.2 Hz, 2H, OCH$_2$), 4.05 (s, 2H, ArCH$_2$), 1.28 (t, J = 7.2 Hz, 3H, OCH$_2$CH$_3$). $^{13}$C NMR (151 MHz, CDCl$_3$) δ= 166.69, 154.38, 154.17, 134.43, 131.11, 130.77, 130.00, 129.52, 127.40, 62.24, 46.79, 30.82, 14.19.

***Tert*-butyl 2-(5-(2-chlorobenzyl)-2-oxo-1,3,4-oxadiazol-3(2*H*)-yl)acetate (7).** The reaction was run with compound **20** (300 mg, 1.42 mmol), DBU (318 μL, 2.13 mmol) and *tert*-butyl bromoacetate (555 mg, 2.84 mmol) in dry THF (6 mL). The crude product was purified by silica gel chromatography (petroleum ether/ethyl acetate 95:5 to 90:10) to give **7** (340 mg, 74.0%) as a white solid. Rf= 0.23 (petroleum ether/ethyl acetate 9:1). MS (ESI$^+$): *m/z* 325 / 327 [M+H]$^+$; $^1$H NMR (600 MHz, CDCl$_3$) δ= 7.40 – 7.39 (m, 1H, ArH$^6$), 7.31 – 7.30 (m, 1H, ArH$^3$), 7.26 -7.24 (m, 2H, ArH$^{4,6}$), 4.32 (s, 2H, NCH$_2$), 4.03 (s, 2H, CH$_2$), 1.44 (s, 9H, CH$_3$). $^{13}$C NMR (151 MHz, CDCl$_3$) δ= 165.63, 154.23, 154.18, 134.36, 131.06, 130.80, 129.94, 129.44, 127.35, 83.39, 47.40, 30.75, 28.06.

**2-(5-(2-Chlorobenzyl)-2-oxo-1,3,4-oxadiazol-3(2*H*)-yl)acetonitrile (21).** The reaction was run with compound **20** (500 mg, 1.42 mmol), DBU (390 μL, 2.61 mmol) and 2-bromoacetonitrile (248 μL, 3.56 mmol) in dry THF (6 mL). The crude product was purified by silica gel chromatography (petroleum ether/ethyl acetate 8:2) to give **21** (540 mg, 91.2%) as a white solid. Rf= 0.45 (petroleum ether/ethyl acetate 7:3). MS (ESI$^+$): *m/z* 250 / 252 [M+H]$^+$; $^1$H NMR (600 MHz, CDCl$_3$) δ= 7.42 – 7.39 (m, 1H, ArH$^6$), 7.34 – 7.29 (m, 1H, ArH$^3$), 7.27 (m, 1H, ArH$^4$), 7.25 (m, 2H, ArH$^5$), 4.33 (s, 2H, NCH$_2$), 4.04 (s, 2H, CH$_2$). $^{13}$C NMR (151 MHz, CDCl$_3$) δ 155.5, 152.6, 134.4, 131.3, 130.1, 130.0, 129.8, 127.5, 112.7, 33.8, 30.8.

**Ethyl 4-((5-(2-Chlorobenzyl)-2-oxo-1,3,4-oxadiazol-3(2*H*)-yl)methyl)benzoate (9).** The reaction was run with compound **20** (300 mg, 1.42 mmol), DBU (318 μL, 2.13 mmol) and ethyl (2-bromomethyl)benzoate (690 mg, 2.84 mmol) in dry THF (6 mL). The crude product was purified by silica gel chromatography (petroleum ether/ethyl acetate 95:5) to give **9** (340 mg, 64.3%) as a white solid. Rf= 0.22 (petroleum ether/ethyl acetate 9:1). MS (ESI$^+$): *m/z* 373 / 375 [M+H]$^+$; $^1$H NMR (600 MHz, CDCl$_3$) δ=8.06 – 7.21 (m, 8H, ArH), 4.88 (s, 2H, NCH$_2$), 4.37 (q, J = 7.1 Hz, 2H, OCH$_2$), 4.00 (s, 2H, CH$_2$), 1.39 (t, J = 7.2 Hz, 3H,

OCH$_2$CH$_3$). $^{13}$C NMR (151 MHz, CDCl$_3$) δ= 166.21, 154.18, 153.89, 139.70, 134.37, 131.10, 130.76, 130.57, 130.17, 129.98, 129.47, 128.05, 127.34, 53.54, 49.12, 30.85, 14.40.

**5-(2-Chlorobenzyl)-3-(4-methoxybenzyl)-1,3,4-oxadiazol-2(3H)-one (10).** To a stirred solution of **20** (200 mg, 0.95 mmol) and 4-methoxy benzyl alcohol (118 mg, 0.95 mmol) in dry THF (5,0 mL), PPh$_3$ (388 mg, 1.42 mmol) was added and the mixture was kept under stirring to 0 °C in a nitrogen atmosphere. A solution of DIAD (279 μL, 1.42 mmol) in dry THF (2,0 mL) was slowly added to the reaction mixture at 0 °C. The mixture was stirred for 4 h and the solvent was removed under reduced pressure. The crude product was purified by silica gel chromatography (petroleum ether/ethyl acetate 95:5 to 90:10) to give **10** (250 mg, 79.9%) as a white solid. Rf= 0.17 (petroleum ether/ethyl acetate 9:1). MS(ESI$^+$) *m/z:* 353 / 355 [M+Na]$^+$. $^1$H NMR (600 MHz, CDCl$_3$) δ= 7.50 – 6.76 (m, 8H, ArH), 4.76 (s, 2H, NCH$_2$), 3.98 (s, 2H, CH$_2$), 3.79 (s, 3H, OCH$_3$). $^{13}$C NMR (151 MHz, CDCl$_3$) δ= 159.74, 153.94, 153.82, 134.40, 131.07, 131.02, 129.98, 129.85, 129.40, 127.33, 127.19, 114.29, 55.41, 49.13, 30.82.

**2-(5-(2-Chlorobenzyl)-2-oxo-1,3,4-oxadiazol-3(2H)-yl)acetic acid (6).** Compound **7** (50 mg, 0.154 mmol) was dissolved in DCM (1.2 mL), trifluoroacetic acid (117 μL, 1.54 mmol) was added and the reaction mixture was stirred at rt overnight. The mixture was diluted with water (15 mL) and extracted with DCM (3 × 15 mL). The organic phases were washed with brine (15 mL), dried (Na$_2$SO$_4$) and concentrated under reduced pressure. The crude product was purified by silica gel chromatography (DCM/MeOH 95:5) to give **6** (27 mg, 63.3%) as a white solid. MS(ESI$^-$) *m/z:* 267 / 269 [M-H]$^-$. $^1$H NMR (600 MHz, DMSO-D$_6$) δ= 7.47-7.42 (m, 2H, ArH), 7.34-7.32 (m, 2H, ArH), 4.43 (s, 2H, NCH$_2$), 4.10 (s, 2H, CH$_2$). $^{13}$C NMR (151 MHz, DMSO-D$_6$) δ= 169.0, 154.3, 154.1, 134.0, 132.3, 131.7, 130.2, 130.1, 130.1, 128.1, 47.1, 30.4.

**2-(5-(2-Chlorobenzyl)-2-oxo-1,3,4-oxadiazol-3(2H)-yl)methyltetrazole (8).** Compound **21** (400 mg, 1.60 mmol) was added to a solution of sodium azide (156 mg, 2.40 mmol) and ammonium chloride (86 mg, 1.60 mmol) in DMF (5 mL) and the reaction mixture was stirred at rt for 18 h. The mixture was diluted with water (15 mL) and extracted with DCM (3 × 15 mL). The combined organic phases were washed with brine (15 mL), dried (Na$_2$SO$_4$) and the solvent evaporated under reduced pressure. The crude product was purified by silica gel chromatography (DCM/MeOH 95:5) to give **8** as a white solid (145 mg, 91%). MS (ESI$^+$) *m/z:* 293 / 295 [M+H]$^+$. $^1$H NMR (600 MHz, CDCl$_3$) δ= 7.97 (s, 1H, NH), 7.36-7.20 (m, 4H, ArH), 5.24 (s, 2H, NCH$_2$), 3.99 (s, 2H, CH$_2$). $^{13}$C NMR (151 MHz, CDCl$_3$) δ= 163.3, 154.9, 153.7, 134.3, 131.2, 130.5, 129.9, 129.5, 127.4, 39.8, 37.0.

**1-(2-Chlorobenzoyl)pyrrolidine-2,5-dione (23).** To a stirred solution of 2-chloro benzoic acid (500 mg, 3.19 mmol) in dry THF (10 mL) kept at 0 °C, N-hydroxysuccinimide (551 mg, 4.79 mmol) and *N,N'*-dicyclohexylcarbodiimide (658 mg, 3.19 mmol). When the addition was complete the ice bath was removed and the reaction was stirred at rt overnight. The obtained suspension was filtered and the solvent removed under reduced pressure. The obtained residue was treated with water (50 mL) and extracted with ethyl acetate (3 × 50 mL). The combined organic phases were washed with brine (150 mL), dried (Na$_2$SO$_4$) and concentrated under reduced pressure. The crude product was purified by silica gel chromatography

(petroleum ether/ethyl acetate 8:2) to give **23** (500 mg, 66.1%) as a white opalescent solid. Rf=0.15 (petroleum ether/ ethyl acetate 8:2). $^1$H NMR (600 MHz, CDCl$_3$) δ= 8.14 – 8.06 (m, 1H, ArH$^6$), 7.59 – 7.50 (m, 2H, ArH$^{3,4}$), 7.42 – 7.37 (m, 1H, ArH$^5$), 2.91 (m, br, 4H, CH$_2$CH$_2$). $^{13}$C-NMR (151 MHz, CDCl$_3$) δ 169.2, 160.2, 135.7, 134.8, 132.5, 131.7, 127.0, 124.6, 25.8.

**Ethyl 1-(5-amino-1,3,4-thiadiazol-2-yl)piperidine-3-carboxylate (25).** 2-amino-5-bromo1,3,4-thiadiazole (300 mg, 1.67 mmol), ethyl nipecotate (284 μL, 1.83 mmol) and DIPEA (1.42 mL, 8.35 mmol) were dissolved in DMF (6 mL) and the mixture was stirred at 80 °C for 1 h. The solvent was evaporated under reduced pressure and the obtained residue treated with water (30 mL) and extracted with ethyl acetate (3 × 30 mL). The combined organic phases were washed with brine (90 mL), dried (Na$_2$SO$_4$) and concentrated under reduced pressure. The crude product was purified by silica gel chromatography (DCM/MeOH 98:2 to 97:3) to give **25** (318 mg, 74.5%) as a white solid. Rf= 0.12 (petroleum ether/ethyl acetate 7:3). MS (ESI$^+$) *m/z*: 257 / 259 [M+H]$^+$. $^1$H NMR (600 MHz, CDCl$_3$) δ= 5.08 (s, 2H, NH$_2$), 4.11 (q, J = 7.1 Hz, 2H, OCH$_2$), 3.83 (m, 1H, PipH$^2$), 3.56 (m, 1H, PipH$^2$), 3.18 (m, 1H, PipH$^6$), 3.11 – 3.03 (m, 1H, PipH$^6$), 2.66 – 2.55 (m, 1H, PipH$^3$), 2.08 – 1.97 (m, 1H, PipH$^4$), 1.75 (m, 1H, PipH$^4$), 1.70 – 1.57 (m, 2H, PipH$^5$), 1.23 (t, J = 7.2 Hz, 3H, CH$_3$). $^{13}$C NMR (151 MHz, CDCl$_3$) δ= 173.11, 165.69, 160.53, 60.83, 51.89, 50.39, 40.69, 26.96, 23.59, 14.28.

**Ethyl *N*-(5-amino-1,3,4-thiadiazol-2-yl)-*N*-methylglycinate (26).** 2-amino-5-bromo1,3,4-thiadiazole (300 mg, 1.67 mmol), ethyl N-methylglicinate hydrochloride (281 mg, 1.83 mmol) and DIPEA (1.08 mL, 8.35 mmol) were dissolved in DMF (7 mL) and stirred at 80°C for 1 h. the solvent was evaporated under reduced pressure and the obtained residue was treated with water (30 mL) and extracted with ethyl acetate (3 × 30 mL). The combined organic phases were washed with brine (90 mL), dried (Na$_2$SO$_4$) and concentrated under reduced pressure. The crude product was purified by silica gel chromatography (DCM/MeOH 98:2) to give **26** (416 mg, 80.4%) as a white solid. Rf= 0.33 (DCM/MeOH 95:5). MS(ESI$^+$) *m/z*: 239 [M+Na]$^+$. $^1$H NMR (600 MHz, DMSO-D$_6$) δ= 6.41 (s, 2H, NH$_2$), 4.10 (s, 2H, NCH$_2$), 4.11-4.07 (m, 3H, OCH$_2$+CH), 3.13 (s, 1H, CH$_3$), 1.21-1.23 (m, 2H), 1.18 (t, J = 7.1 Hz, 3H, CH$_3$). $^{13}$C NMR (151 MHz, DMSO-D$_6$) δ= 169.9, 162.9, 161.0, 61.0, 54.1, 42.4, 18.6, 17.3, 14.6, 13.0.

**Ethyl *N*-(5-amino-1,3,4-thiadiazol-2-yl)-*N*-benzylglycinate (27).** 2-amino-5-bromo1,3,4-thiadiazole (500 mg, 2.78 mmol), ethyl N-benzylglicinate (573 μL, 3.06 mmol) and DIPEA (0.92 mL, 5.55 mmol) were dissolved in DMF (6 mL) and stirred at 80°C for 1 h. The solvent was evaporated under reduced pressure and the obtained residue was diluted with water (30 mL) and extracted with ethyl acetate (3 × 30 mL). The combined organic phases were washed with brine (90 mL), dried (Na$_2$SO$_4$) and concentrated under reduced pressure. The crude product was purified by silica gel chromatography (petroleum ether/ethyl acetate/methanol 7:2.5:0.5) to give **27** (712 mg, 77.6%) as a white solid. Rf= 0.12 (petroleum ether/ethyl acetate 7:3). MS (ESI$^+$): *m/z* 293 [M+H]$^+$; $^1$H NMR (600 MHz, CD$_3$CN) δ= 7.89 (s, 1H, ArH), 7.34-7.27 (m, 4H, ArH), 5.34 (s, 2H, NH$_2$), 4.53 (s, 2H, CH$_2$), 4.10 (m, 4H, 2 x CH$_2$), 1.17 (t, J = 7.2 Hz, 3H). $^{13}$C NMR (151 MHz, CD$_3$CN) δ= 169.6, 162.5, 160.4, 136.9, 128.7, 127.9, 127.8, 57.3, 51.9, 35.7, 13.6.

**General procedure for the synthesis of compounds 11-13.** The appropriate 5-amino-1,3,4-thoadiazole intermediate **25-27** was dissolved in DMF, then an equimolar amount of the activated ester **23** and DIPEA were added. The reaction mixture was stirred at 100 °C overnight. The reaction mixture was treated with a saturated solution of NaHCO$_3$ (30 mL) and extracted with ethyl acetate (3 × 30 mL). The combined organic phases were washed with brine (90 mL), dried (Na$_2$SO$_4$) and concentrated under reduced pressure. The crude product was purified by silica gel chromatography as described below.

**Ethyl 1-(5-(2-chlorobenzamido)-1,3,4-thiadiazol-2-yl)piperidine-3-carboxylate (11).** The reaction was run with **25** (100 mg, 0.390 mmol), **23** (92.4 mg, 0.390 mmol) and DIPEA (66 μL, 0.390 mmol). After standard work-up as described above, the crude product was purified by silica gel chromatography (petroleum ether/ethyl acetate 7:3 to 6:4) to give **11** (130 mg, 84.9%) as a white solid. Rf= 0.18 (DCM/MeOH 99:1). MS (ESI$^+$) *m/z* 395 / 397 [M+H]$^+$. $^1$H NMR (600 MHz, CDCl$_3$) δ=7.78 (dd, J = 7.4, 1.7 Hz, 1H, ArH$^6$), 7.41 (m, 3H, ArH$^{3,4,5}$), 4.18 (q, J = 7.1 Hz, 2H, OCH$_2$), 3.96 (m, 1H, PipH$^2$), 3.63 (m, 1H, PipH$^2$), 3.19 (m, PipH$^6$), 3.13 – 3.07 (m, 1H, PipH$^6$), 2.60 (m, 1H, PipH$^3$), 2.15 – 2.07 (m, 1H, PipH$^4$), 1.78 (m, 1H, PipH$^4$), 1.74 – 1.59 (m, 2H, PipH$^5$), 1.29 (t, J= 7.1 Hz, 3H, CH$_3$). $^{13}$C NMR (151 MHz, CDCl$_3$) δ= 172.97, 167.78, 164.15, 151.86, 132.80, 132.14, 130.65, 130.59, 127.15, 60.90, 50.94, 50.07, 40.55, 27.10, 23.65, 14.36.

**Ethyl *N*-methyl-*N*-(5-(2-chlorobenzamido)-1,3,4-thiadiazol-2-yl)glycinate (12).** The reaction was run with **26** (250 mg, 0.975 mmol), **23** (231 mg, 0.975 mmol), and DIPEA (165 μL, 0.975 mmol). The crude product was purified by silica gel chromatography (petroleum ether/ethyl acetate 7:3 to 6:4) to give **12** (325 mg, 81.9%) as a white solid. Rf= 0.18 (DCM/MeOH 99:1). MS(ESI$^+$) *m/z* 377 / 379 [M+Na]$^+$. $^1$H NMR (600 MHz, DMSO-D$_6$) δ= 12.67 (s, 1H, NH), 7.63-7.44 (m, 4H, ArH), 4.33-4.30 (m, 2H, OCH$_2$), 4.15-4.12 (m, 2H, NCH$_2$), 3.15-3.12 (m, 3H, NCH$_3$), 1.21 (t, J = 7.2 Hz, 3H, CH$_3$). $^{13}$C NMR (151 MHz, DMSO-D$_6$) δ= 176.9, 176.5, 169.2, 134.2, 132.0, 130.3, 129.8, 129.5, 127.2, 60.7, 53.2, 40.4, 14.1.

**Ethyl *N*-benzyl-*N*-(5-(2-chlorobenzamido)-1,3,4-thiadiazol-2-yl)glycinate (13).** The reaction was run with **27** (500 mg, 2.61 mmol), **23** (408 mg, 2.61 mmol), and DIPEA (47 μL, 2.61 mmol). The crude product was purified by silica gel chromatography (petroleum ether/ethyl acetate 7:3 to 6:4) to give **13** (618 mg, 82.3%) as a white solid. Rf= 0.41 (petroleum ether/ethyl acetate/methanol 7:3). MS(ESI$^+$) *m/z* 431 / 433 [M+H]$^+$. $^1$H NMR (600 MHz, CDCl$_3$) δ= 7.77 (d, J = 7.6 Hz, 1H, ArH), 7.42 (d, J = 7.9 Hz, 1H, ArH), 7.39-7.31 (m, 5H, ArH), 7.28 (d, J = 7.2 Hz, 2H, ArH), 4.65 (s, 2H, NCH$_2$), 4.20 (q, J = 7.1 Hz, 2H, OCH$_2$), 4.07 (s, 2H, NCH$_2$), 1.27 (t, J = 7.1 Hz, 3H, CH$_3$). $^{13}$C NMR (151 MHz, CDCl$_3$) δ= 169.0, 167.3, 164.0, 151.4, 135.3, 132.6, 132.1, 132.0, 130.5, 130.5, 128.9, 128.2, 127.9, 127.0, 61.4, 56.8, 50.7, 14.2.

*4.3. Biological Evaluations*

**Cell culture and treatment.** Human monocytic THP-1 cells were cultured in RPMI 1640 medium (Aurogene, Rome, Italy) supplemented with fetal bovine serum (10%; Aurogene), L-glutamine (2 mM; Aurogene), penicillin (100 IU/mL; Aurogene) and streptomycin (100 mg/mL; Aurogene). Cell culture medium was replaced every 2−3 days, and the cultures were maintained at 37 °C and 5% CO2 in a fully humidified incubator. The day before each experiment, THP-1 cells were plated in 48-well plates (90.000 cells/well) and differentiated into macrophages by treatment with phorbol myristate acetate (PMA; 50 nM; 24 h; Sigma-Aldrich). Differentiated cells were washed twice with phosphate-buffered saline (PBS) and then were primed with lipopolysaccharide (LPS; 10 μg/mL; 4 h; Sigma-Aldrich) in serum-free medium. Cells were treated with either vehicle alone or test compound (10 μM; 1 h) and cell death was triggered with ATP (5 mM, 90 minutes, Sigma-Aldrich). In dose-response experiments, cells were treated with different compound concentrations in the range 0.1-100 μM.

**Pyroptotic cell death.** The pyroptotic cell death was quantified by determining LDH activity using the Cytotox 96 nonradioactive cytotoxicity assay (Promega Corporation, Madison, MI, USA). Absorbance was measured using the Victor X4 (PerkinElmer, Waltham, MA, USA) at λ=490nm. Cell death was expressed according to the manufacture's instruction.

**Il-1β and TNF-α release.** IL-1β and TNF-α release was quantified in differentiated THP-1 supernatant, obtained as previously described, using Human IL-1β Uncoated ELISA kit and Human TNF-α Uncoated ELISA kit respectively (Invitrogen, Waltham, MA, USA), according to the manufacture's instruction.

**Cytotoxicity Assay.** THP-1 were plated in 96-wells culture plates (15.000 cells/well) and then treated with increasing concentrations (0.1–100 μM) of each compound. Cell viability was measured at 72 h by the MTT assay, a colorimetric assay based on the conversion of the water-soluble 3-(4,5-dimethylthiazol-2-yl)- 2,5-diphenyltetrazolium bromide (MTT; Sigma-Aldrich) to an insoluble purple formazan by actively respiring cells. The formazan concentration was determined by measuring absorbance in the Victor X4 at a λ = 570 nm.

**THP-1 cells lysate.** THP-1 cells were plated in 10 cm plates (1.500.000 cells/plate) and differentiated into macrophages as previously described. After 24h, cells were washed twice with ice-cold PBS and lysed using RIPA buffer.

**Statistics.** Experiments were performed at least three times. Graphs were constructed, data analyzed, and $IC_{50}$ for each compound calculated using GraphPad Prism 9. Dose–response curves were analyzed using a log (agonist) vs. normalized response model. Where indicated, the results are given as the mean ± S.E.M. Statistical analyses were performed by one-way ANOVA using Dunnett's post hoc test (GraphPad).

*4.4. Computational studies*

**Modelling.** The X-ray crystal structure of NLRP3 NACHT domain (residues 134-676, PDB ID: 7ALV, resolution = 2.83 Å) was selected as the target protein. For modelling missing side chains and loops distant from the active site (residues 153-163, 178-200, 213-216, 452-462, 496-497, 513-515,539-554, 589), Uniprot sequence of NLRP3 (Q96P20) was modelled on PDB ID structure 7ALV. Four Ramachandran outliers

deriving from the model (194 LYS, 199 PRO, 540 LYS, 551 LEU) were in the modelled loops, distant from the active site (see Figure S12-14).

**Docking.** Compounds were docked using the Maestro suite by Schrödinger, v. 2022.1. Protein was prepared with the Protein Preparation Wizard tool, at PROPKA pH = 7.4, heavy atoms were minimized with restraints to RMSD 0.3 Å with OPLS 4 force field. Ligands were first prepared using the Ligprep module, generating possible states at pH = 7.4. After self-docking of the NP3-146 ligand in the binding site (Figure S15), each candidate ligand was docked using the Induced-Fit Docking (IFD) extended protocol [85], generating up to 80 poses, and further refining residues Arg351 and Arg578. For the oxadiazole series, residues Thr439, Met408, Ile411 and Phe575 were not allowed to move for preserving the shape of the lipophilic pocket, while for all the other ligands such constraint was not necessary. The binding site was defined as the centroid of residues Ala228, Arg351 and Arg 578, the grid was enlarged to host ligands with length up to 30 Å. Compounds were docked in their putative active form, i.e.: hydrolyzed ester group, with a free carboxylic acid. The ten top-ranked poses according to the IFD Score were visually inspected for assessing the preferred binding mode.

**Ligand parameterization.** Compound **6** and NP3-146 ligands were parameterized using BiKiLifeSciences version 1.5. Atomic partial charges were assigned using the Resp-DFT protocol as implemented in the Residue Parameterization tool in BiKi [www.bikitech.com].

**MD setup and analysis.** The ionization state of histidine residues at pH = 7.4 was checked via the H++ webserver (http://newbiophysics.cs.vt.edu/H++/): as such, three histidine residues were found charged (HIP215, HIP220 and HIP364), three histidine residues were protonated on $N_\delta$ (HID260, HID492 and HID663), while all the other histidine residues were protonated on $N_\epsilon$. The *a99SB-disp* forcefield imported in GROMACS 2022.1 was used to parameterize the protein, and related modified TIP4P-D parameters were used for water molecules [86]. The system was embedded in a truncated dodecahedral box and solvated with modified TIP4P-D water contained in *a99SB-disp* forcefield itself. Charges were neutralized using NaCl at 0.15 M concentration. Periodic boundary conditions (PBC) were set, long range electrostatic effects were adjusted with the Particle-Mesh Ewald (PME) method, a cut-off value of 12 Å was fixed for both electrostatic and van der Waals interactions. The systems underwent $1 \cdot 10^6$ energy minimization cycles using the steepest descendant algorithm, followed by a further refinement using the conjugate gradient algorithm. The integration timestep was set to 2 fs, MD frames were saved every 10 ps. The temperature was gradually increased in six steps in the canonical ensemble (NVT), from 0 to 300 K, rescaling velocities to control the system temperature. Position restraints were applied on protein and ligands during the first four thermalization steps (from 0 to 200 K) and were then removed in the last two steps (200 to 300 K). Two consecutive equilibration steps (5 ns) were run in the isothermal-isobaric ensemble (NPT) using the C-rescale barostat, with only the first applying position restraints to protein and ligands. Finally, production in the NVT ensemble was run for 300 ns. The MD protocol was replicated five times for each system, i.e., NLRP3+NP3-146 and NLRP3+Compound **6**.

Analysis of MD simulations was carried out using GROMACS 2022.1 built-in functions. RMSD and RMSF were calculated with *gmx_rms* and *gmx_rmsf* tools. Hydrogen bonds were evaluated with the *g_hbond* tool, using both distance (< 0.35 nm) and angle (angle < 30°) cutoffs. In particular, the hydrogen bond angle is the measured angle between the hydrogen atom H and the acceptor atom A, measured at the donor atom D (∡HDA), while the cut-off distance is calculated between the donor and the acceptor atoms. Hydration of binding sites was calculated using the Hydra tool in BiKiLifeSciences version 1.5. Plots were generated using Matplotlib v. 3.5.2 and Seaborn v. 0.11.2, figures were obtained using PyMOL v. 2.5.2.

*4.5. Stability studies*

**Stability in PBS.** A solution of compound (10 mM in DMSO) was added to phosphate buffer saline (PBS, 50 mM) (2.3 g of disodium hydrogen orthophosphate, 0.19 g of potassium dihydrogen orthophosphate and 8.0 g of sodium chloride in 1000 mL, pH adjusted to 7.4). The resulting solutions (100 μM) were kept at 37 ±0.5°C for 48 hours; aliquots of 20 μL were withdrawn at appropriate time intervals and analyzed by RP-HPLC (as reported below). Each experiment was independently repeated at least three times. The results are expressed as % of unmodified compound and % of degradation product during the 48 hours incubation.

**Stability in human serum.** A solution of compound (10 mM in DMSO) was added to human serum (sterile-filtered from human male AB plasma, USA origin, sterile-filtered, Sigma–Aldrich) preheated and maintained at 37 ±0.5°C. 200 μL of the resulting solution (100 μM) was withdrawn at appropriate time intervals and added to 200 μL of $CH_3CN$ containing 0.1% TFA in order to deproteinize the serum. Sample was sonicated, vortexed and then centrifuged for 10 minutes at 2150 *g*. The clear supernatant was filtered by PTFE filters (25mm 0.45μm) and analyzed by RP-HPLC as described below. The results are expressed as % of unmodified compound and % of degradation product during the 24 hours incubation. Pseudo-first order half-times ($t_{1/2}$) for the hydrolysis was calculated with exponential decay equation model by Graph-Pad Prism v. 7.0 (GraphPad Software Inc., San Diego, CA).

**Stability in THP-1 cells lysate**. A solution of compound (10 mM in DMSO) was added to THP-1 cells lysate preheated and maintained at 37 ±0.5°C. 200 μL of the resulting solutions (100 μM) was withdrawn at appropriate time intervals and added to 200 μL of $CH_3CN$ containing 0.1% TFA in order to deproteinize the sample. Sample was sonicated, vortexed and then centrifuged for 10 minutes at 2150 *g*. The clear supernatant was filtered by PTFE filters (25mm 0.45μm) and analyzed by RP-HPLC as described below. The results are expressed as % of unmodified compound and % of degradation product during the 24 hours incubation. Pseudo-first order half-times ($t_{1/2}$) for the hydrolysis was calculated with exponential decay equation model by Graph-Pad Prism v. 7.0 (GraphPad Software Inc., San Diego, CA).

**Stability in vehicle for intraperitoneal administration**. Compound **5** (INF200) solution (2 mg/mL) in saline, 2% DMSO, 5% Cremophor-EL was mantenied for 7 days at 4° C and 25°C, and analyzed every day by RP-HPLC as described below.

**HPLC analysis.** HPLC analyses of stability studies were performed with a HP1200 chromatograph system (Agilent Technologies, Palo Alto, CA, USA) equipped with a quaternary pump (model G1311A), a membrane degasser (G1322A), a multiple wavelength UV detector (MWD, model G1365D) integrated in the HP1200 system. Data analysis was performed using a HP ChemStation system (Agilent Technologies). The sample was eluted on a ZORBAX SB-Phenyl column (250 × 4.6 mm, 5 μm, Agilent Technologies, Palo Alto, CA, USA). The injection volume was 20 μL (Rheodyne, Cotati, CA). The mobile phase consisting of acetonitrile 0.1% TFA and 0.1% TFA 55/45 %v/v at flow-rate = 1.0 mL/min. The column effluent was monitored at 226 and 254 nm referenced against a 800 nm wavelength. Data analysis was performed with Agilent ChemStation. Quantitation of compounds was done using calibration curves of compounds obtained analysing standard solutions in a concentration range of 1-100 μM ($r^2 > 0.99$).

*4.6. Determination of the lipophilic hydrophilic balance*

The distribution coefficients (log D) between n-octanol and water were obtained by shake-flask technique at room temperature. In the shake-flask experiments phosphate 50 mM buffers (pH 7.4) was used as aqueous phases; ionic strength was adjusted to 0.15 M with KCl. The organic (*n*-octanol) and aqueous phases were mutually saturated by shaking for 4 h. The compounds were solubilised in the buffered aqueous phase at a concentration of about 0.1 mg/ml and appropriate amounts of *n*-octanol were added. The two phases were shaken for about 20 min, by which time the partitioning equilibrium of solutes is reached, and then centrifuged (10000 rpm, 10 min). The concentration of the solutes was measured in the aqueous phase by UV spectophotometer (UV-2501PC, Shimadzu). Each log $D^{7.4}$ value is an average of at least six measurements.

*5. **In vivo** study*

*5.1. Drugs and reagents*

KCl, NaCl, NaHCO$_3$, CaCl$_2$, MgSO$_4$, KH$_2$PO$_4$, NaH$_2$PO$_4$, mannitol, glucose, Na-pyruvate, β-nicotinamide adenine dinucleotide (NADH), thiobarbituric acid (TBA), bovine serum albumin (BSA), butanol-1, butylated hydroxyanisole, diethyl ether, diethylenetriamine pentaacetic acid, ethylenediaminetetraacetic acid (disodium salt), tween-20 and pyrogallol were purchased from Sigma Aldrich (St. Louis, MO, USA). Absolute ethanol, hydrochloric acid, methanol, and trichloroacetic acid were purchased from Carlo Erba Reagents (Cornaredo, MI, Italy). All the solutions were freshly prepared before starting the experiments.

*5.2. Animals*

This study was conducted on male Wistar rats (Envigo-Udine, Italy), individually housed in cages under controlled light (12 h light/dark cycle) and temperature (23–25 °C; 50–55% humidity) conditions and fed *ad libitum*. Experiments were performed in accordance with the Declaration of Helsinki, the Italian law (D.L. 26/2014), the Guide for the Care and Use of Laboratory Animals published by the US National Institutes of Health (2011) and the Directive 2010/63/EU of the European Parliament on the protection of animals used for science. The project was approved by the Italian Ministry of Health, Rome, and by the ethics review board.

*5.2.1. Experimental groups:*

Animals were divided in four experimental groups fed with Standard Diet (Diet 2018, used as control diet, and indicated as SD; 6.2% kcal fat, 18.6% kcal protein, and 44.2% kcal carbohydrate, provided by Envigo, Udine, Italy) or High Fat Diet (Teklad Diet TD 06414, used as hypercaloric diet, and indicated as HFD; 60% kcal fat, 18.3%, kcal protein and 21.7% kcal carbohydrate, provided by Envigo, Udine, Italy) for 12 weeks, as described in previous publications [46-48]. Rats were randomly assigned to the different experimental groups and in the last 28 days of the diet protocols they were treated with intraperitoneal (i.p.) injection of the selective inhibitor of NLRP3 (INF200) (20 mg/kg/day) or vehicle solution [87, 88]. In particular, the experimental groups were divided as follow: (1) *Group SD* (n=6, fed with SD and i.p. treated with vehicle in the last 28 days of the SD protocol); (2) *Group SD + INF200* (n=7, fed with SD and i.p. treated with INF200 20 mg/kg/day in the last 28 days of the SD protocol); (3) *Group HFD* (n=6, fed with HFD and i.p. treated with vehicle in the last 28 days of the HFD protocol); (4) *Group HFD + INF200* (n=7, fed with HFD and i.p. treated with INF200 20 mg/kg/day in the last 28 days of the HFD protocol).

*5.3. Anthropometric variables*

Body weight was measured weekly, while abdominal, epididymal, retroperitoneal and perirenal fat and liver were removed and weighed individually after sacrifice. Heart weight was determined to calculate the cardiac somatic index (CSI), as the ratio of heart weight to body weight multiplied by 100, as previously described [46-48].

*5.4. Plasma biochemical analysis*

*5.4.1. Metabolic parameters*

After sacrifice, the serum was obtained by centrifuging the blood at 2000 x *g* for 10 min at 4° C. Glycaemia was determinate by using a glucometer (ACCUCHEK, Roche Diagnostics, Germany); total cholesterol, HDL cholesterol, LDL cholesterol and triglycerides were determined using PKL® POKLER ITALY kits as previously described [46-48].

*5.4.2. Enzyme-linked immunosorbent assay (ELISA)*

Detection of plasma interleukin-1β (IL-1β), tumor necrosis factor-α (TNF-α), brain natriuretic peptide (BNP) (Elabscience Biotechnology Inc. United Sates), galectin-3 (Gal-3) (MyBioSource, Inc. United Sates) were carried out by specific ELISA kits according to the manufacturer's instructions.

*5.4.3. Lactate dehydrogenase (LDH) determination*

LDH enzymatic activity was measured in the plasma samples of each experimental groups following the method of McQueen (1972) and as previously reported [78, 81].

**6. Ex vivo study**

At the end of the *in vivo* experimental protocols above described, rats were anesthetized with i.p. injections of ethyl carbamate (2 g/kg body weight) and subsequently euthanized. The hearts were rapidly excised, placed in ice-cold perfusion buffer, cannulated through the aorta and perfused according to the Langendorff method at a constant flow of 12 ml/min (37°C), as reported in previous publications [74, 75, 81]. Perfusion was performed by using Krebs – Henseleit (KH) buffer containing 4.7 mM KCl, 113 mM NaCl, 25 mM $NaHCO_3$, 1.8 mM $CaCl_2$, 1.2 mM $MgSO_4$, 1.2 mM $KH_2PO_4$, 1.1 mM mannitol, 11 mM glucose and 5 mM Na-pyruvate (pH 7.4; 37 °C; 95% $O_2$ and 5% $CO_2$). A latex balloon, filled with water, connected to a pressure transducer (BLPR; WRI, Sarasota, FL, USA), was placed in the LV through the mitral valve, to record cardiac mechanical parameters. The balloon was filled to obtain a LV end-diastolic pressure (LVEDP) of 5–8 mmHg during basal conditions. A second pressure transducer, located above the aorta, was employed for recording the coronary pressure (CP). Hemodynamic parameters were monitored and analyzed using a PowerLab data acquisition system (AD Instruments, Sydney, New SouthWales, Australia) and the performance variables were measured every 10 min.

*6.1. Protocols of Ischemia/Reperfusion injury (IRI)*

At the end of the *in vivo* treatments, the hearts of each experimental group were subjected to IRI protocols. To induce IRI, after stabilization (40 min), the hearts were subjected to 30 min of global, no-flow ischemia followed by 120 min of reperfusion, while the sham hearts were only perfused with KH buffer for 190 min. Cardiac performance was analysed before, during and after ischemic event by monitoring the developed LV pressure (dLVP) recovery, an index of contractility, and LVEDP parameter, an index of contracture (defined as an increase of 4 mmHg above the baseline level), as previously described in the rat heart [75, 78, 81].

*6.3. Assessment of Myocardial Injury*

*6.3.1. Evaluation of infarct size (IS)*

At the end of IRI protocols, the hearts were removed from the Langendorff apparatus, and infarct areas were measured by nitro blue tetrazolium staining by an independent observer and in a blinded manner, as previously reported [75, 78, 81]. After staining, unstained necrotic tissues and stained viable tissues were carefully separated and weighed. The infarct size (IS) was assessed as percentage of the total LV mass, as previously described [75, 78, 81].

*6.3.2. Evaluation of LDH activity in coronary effluents of IRI exposed hearts*

The coronary effluents of hearts subjected to IRI of each experimental group were collected in ice-cold tubes immediately before ischemia and at 10, 20, and 30 min of reperfusion to determine the enzymatic activity of LDH, according to McQueen method (1972) and as previously reported [75, 77, 81]. LDH activity was spectrophotometrically evaluated by a Multiskan™ SkyHigh (Thermo Fisher Scientific Inc., Waltham, MA, USA) by monitoring the absorbance decrease at 340 nm resulting from NADH oxidation and expressed as IU/L.

*6.4. Western Blot and biochemical analyses*

After IRI protocols LV of the hearts of each experimental group were separately processed for immunoblot and biochemical analyses to evaluate specific inflammatory and oxidative stress-related markers.

*6.4.1. Western blot analysis for NLRP3 detection*

LV tissues were homogenized in ice-cold RIPA lysis buffer containing a mixture of protease and phosphatase inhibitors (1 mmol/L aprotinin, 20 mmol/L phenylmethylsulfonyl fluoride, and 200 mmol/L sodium orthovanadate) and centrifuged at 15,000 x *g*, for 25 min, at 4°C. Supernatant was collected and the protein concentration was determined by using Bradford assay (Sigma-Aldrich, St. Louis, MO, USA). 50 micrograms of protein were loaded on 8% sodium dodecyl sulphate–polyacrylamide gel electrophoresis (SDS–PAGE) gels for NLRP3 and transferred by Trans-Blot Turbo system to Nitrocellulose membrane (Bio-Rad Laboratories, Inc.). Then, membranes were blocked with 5% non-fat dry milk in Tris-buffered saline containing 0.1% Tween-20 (TBST) for 1 hour and incubated overnight at 4°C with the primary specific antibody against NLRP3 (NBP2-12446, Novus Biologicals, Milan, Italy) diluted 1:500 in TBST containing 5% Bovine Serum Albumin (BSA). The antibody against β-actin was used as loading control. After incubation with the primary antibodies, the membranes were washed 3 times for 5 min with TBST and incubated with the horseradish peroxidase-conjugated secondary antibodies (in TBST containing 5% non-fat dry milk) for 1 hour at room temperature (Anti-Rabbit or Anti-Mouse, Sigma Aldrich, 1: 3000 and 1:2000, respectively). Immunodetection was performed with enhanced chemiluminescence (ECL) Western blotting detection system (Santa Cruz Biotechnology) and densitometric analyses of bands were carried out by ImageJ 1.6 software (National Institutes of Health, Bethesda, MD, USA), after digitalization by evaluating the areas and the pixel intensity and subtracting the background, as previously reported [78, 81].

*6.4.2. Biochemical analyses*

*6.4.2.1. Enzyme-linked immunosorbent assay (ELISA) for inflammatory cytokines detection*

Detection of cardiac interleukin-1β (IL-1β) and tumor necrosis factor-α (TNF-α) was performed in LV tissues of sham and IRI hearts by using specific ELISA kits, as above reported, according to the manufacturer's instructions.

*6.4.2.2. Thiobarbituric Acid Reactive Substances (TBARS) Assay*

The levels of malondialdehyde (MDA) were detected in LV samples of sham and IRI hearts as a specific indicator of lipid peroxidation. MDA concentration was assessed by thiobarbituric acid reactive substances (TBARS) assay, as previously described [71, 78, 89]. LV samples were homogenized in 0.9% KCl (pH 7.4) (10% w/v), incubated at 37 °C, and 1 ml of 40% (w/v) TCA and 1 ml of 0.2% (w/v) TBA were added to 2 ml of each homogenate. 2% (w/v) butylated hydroxytoluene was also added to the TBA mixture for preventing artificial lipid peroxidation during the assay, as previously indicated [90]. Following incubation at 100 °C for 15 min, 2 ml of 70% (w/v) TCA were added and the mix was allowed to stand for 20-25 min at room temperature. Samples were then centrifuged at 3500 rpm for 20 min and the colorimetric intensity, assessed spectrophotometrically at 523 nm, was expressed as nanomoles MDA/g heart tissue.

*6.4.2.3. Superoxide dismutase (SOD) assay*

SOD activity in sham and LV extracts was evaluated according to the method of Marklund and Marklund (1974) [91] following the inhibition of pyrogallol autoxidation, and as previously described [78]. In brief, LV tissues were homogenized in 50 mM Tris HCl, pH 8.2 containing 1 mM diethylenetriamine pentaacetic acid, centrifuged at 20,000 x *g*, 20 min, and supernatants were collected for SOD determination. After adding 0.2 mM pyrogallol, its autoxidation was monitored by measuring absorbance at 420 nm for 3 min (Multiskan™ SkyHigh, Thermo Fisher Scientific Inc., Waltham, MA, USA). The enzyme activity was assessed as percentage of pyrogallol autoxidation, where 1 U of SOD corresponds to the amount of enzyme that inhibits the rate of pyrogallol autoxidation by 50%.

*7. Statistical Analyses*

Data were expressed as mean ± SEM. A two-way ANOVA and nonparametric Bonferroni's multiple comparison test (for post-ANOVA comparisons) were used for analysing body weight, hemodynamic results and LDH enzymatic activity in coronary effluents. A one-way ANOVA and the nonparametric Newman–Keuls multiple comparison test (for post-ANOVA comparisons) were used for the other analyses. Values (* $p < 0.05$, ** $p < 0.01$, *** $p < 0.001$, **** $p < 0.0001$) were considered statistically significant. Lines denote the comparisons among the experimental groups. The statistical analysis was conducted using Prism 5 (GraphPad Software, La Jolla, CA, USA).

**Funding sources**

This project has received funding from the European Union's Horizon 2020 Framework Programme for Research and Innovation under the Specific Grant Agreement No. 945539 (Human Brain Project SGA3) " Project BRAVE"; University of Turin, Ricerca Locale 2021, 2022 (BERM_RILO_21_01; SPY_RILO_21_01; SPY_RILO_22_01; MARE_RILO_22_01). This research was funded by the National Institute of Cardiovascular Research (INRC), Bologna, Italy, in the context of the research project "The challenge of dealing with heart failure with preserved ejection fraction: multiple phenotypes with a common pathophysiological substratum?", *Donazione AnnaMaria Ruvinetti* and the University of Calabria "ex-60%".

**Declaration of competing interests**

The authors declare no competing financial interests

**Data Availability**

The data presented in this study are contained within the article and its supplementary material.

**Acknowledgements**

We acknowledge the use of Fenix Infrastructure resources, which are partially funded from the European Union's Horizon 2020 research and innovation programme through the ICEI project under the grant agreement No. 800858 (Application No. 25681). We kindly thank BiKi technologies for providing the BiKi Life Science suite. Maria Concetta Granieri (M.C.G.) acknowledges "Programma Horizon Europe", host institution University of Calabria (DR. n. 1101 del 29/07/2022), for the Post-doc position. Simone Gastaldi and Eleonora Ginaquinto acknowledge Horizon 2020 Framework Programme, Specific Grant Agreement No. 945539 (Human Brain Project SGA3) for the post-doc positions.

**Author Contributions**

Conceptualization: Carmine Rocca, Pasquale Pagliaro, Claudia Penna, Tommaso Angelone, Francesca Spyrakis, Massimo Bertinaria
Methodology: Carmine Rocca, Maria Concetta Granieri, Anna De Bartolo, Naomi Romeo, Rosa Mazza, Simone Gastaldi, Eleonora Gianquinto, Margherita Gallicchio, Barbara Rolando, Elisabetta Marini.
Validation: Carmine Rocca, Maria Concetta Granieri, Anna De Bartolo, Naomi Romeo, Rosa Mazza, Francesco Fedele, Margherita Gallicchio, Pasquale Pagliaro, Claudia Penna, Tommaso Angelone, Francesca Spyrakis, Massimo Bertinaria
Investigation: Carmine Rocca, Maria Concetta Granieri, Anna De Bartolo, Naomi Romeo, Rosa Mazza, Francesco Fedele, Eleoonora Gianquinto, Simone Gastaldi, Federica Blua, Valentina Boscaro, Pasquale Pagliaro, Claudia Penna, Tommaso Angelone, Francesca Spyrakis, Massimo Bertinaria
writing—original draft preparation: Carmine Rocca, Maria Concetta Granieri, Simone Gastaldi, Eleonora Gianquinto, Massimo Bertinaria
writing—review and editing: Carmine Rocca, Francesco Fedele, Pasquale Pagliaro, Claudia Penna, Tommaso Angelone, Francesca Spyrakis, Massimo Bertinaria